\documentclass[12pt]{iopart}
\usepackage{graphicx}
\usepackage{caption}
\usepackage{float}

\usepackage{amssymb}
\usepackage{subcaption}
\usepackage{mwe}
\usepackage{hyperref}
\hypersetup{ colorlinks=true, linkcolor=blue, urlcolor=marineblue, citecolor=blue, filecolor=magenta, urlcolor=cyan,
 }
\begin{document}

\title[Insight into the evolution of laser-induced plasma]{Insight into the evolution of laser-induced plasma during successive deposition of laser energy}

\author[1]{Awanish Pratap Singh}
\address {Institute of Biomedical Optics, University of Lübeck, Peter-Monnik-Weg 4, 23564 Lübeck, Germany}
\ead{awanishsingh009@gmail.com}
\author{Upasana P. Padhi}
\address {Department of Aerospace Engineering, Indian Institute of Technology Bombay, Mumbai, Maharashtra 400076, India}
\ead{upasana0904@gmail.com}
\author { Ratan Joarder}
\address {Department of Aerospace Engineering, Indian Institute of Technology Kharagpur, West Bengal, India, 721302}
\ead{jratan@aero.iitkgp.ac.in}
\vspace{10pt}


\begin{abstract}
The interaction of high-temperature plasma with matter has several potential applications. In this study, we generated laser-induced plasma through single and successive laser energy deposition. 
The lifetime of the plasma is of paramount importance in most practical applications. However, this cannot be achieved with a single high-energy pulse due to some practical challenges. Therefore we carried out experimental and numerical investigations on the successive laser energy deposition and demonstrated its importance compared to the single pulse energy deposition. It has been observed that during successive energy deposition, the absorption of energy from the second pulse is nonlinear, and the reason for such behaviour is explained in this study. Due to the nonlinear absorption from the second pulse, this study therefore aims to present the pulse interval configuration between the successive pulses with which it is effective for practical use. In this study, some interesting and first direct observations of some new physical phenomena (generation of fourth-lobe and multiple shock waves) are observed during successive energy deposition when the pulse interval is 50 $ \mu{s} $ and 100 $ \mu{s} $. This study also adopted new approch based on Maxwell's theory of momentum exchange between light and matter to explain the newly observed and some previously unexplained physical phenomena. Finally, to understand the evolution of the laser-induced plasma, the volume and volumetric expansion rate are calculated, which can be useful in determining its lifetime and mixing rate with the surrounding medium.

\end{abstract}

%
%
%
%
%

\section{Introduction}
\label{S:1}

Laser-induced breakdown (LIB) has attracted significant interest due to the wide range of potential applications. The single pulse LIB (SP-LIB) and the successive laser energy depositions (LED) have been used to generate plasma in various fields of science and engineering, such as ignition and combustion under different conditions \cite{wermer2017dual, bak2015successive}, spectroscopic analyses \cite{gebel2015optical, tognoni2014basic}, aerodynamic drag reduction \cite{joarder2017two, joarder2018mechanism, Joarder_2019b}, lift and moment control \cite{daso2009dynamics}, biology and medicine \cite{Vogel_2003, Vogel_2005, Cherian_2008}, surface cleaning \cite{Seo_2015}, metallurgy and material processing \cite{Cao_2003, Nivas_2015}, local flow-field alternation \cite{schulein2008experimental}, propulsion system and spacecraft thrusters \cite{manfletti2013laser}, etc. Extensive research has been performed to understand the process of energy deposition with SP-LIB and it has been well understood. However, the physical phenomena associated with successive LEDs are quite complex and not yet clearly understood. For successive double-pulse LED (DP-LED), the first-pulse breakdown follows the SP-LIB trend, while the absorption from the second-pulse shows a different behaviour when the pulse interval changes. The change in the absorption behaviour of the second pulse has been observed in several studies \cite{bak2015schlieren, an2017experimental}. However, to the knowledge of the authors, the actual reason for such a change is nowhere given. In previous studies \cite{wermer2017dual, bak2015successive, bak2015schlieren, an2017experimental}, it has been reported that the second pulse in DP-LED causes another breakdown at all pulse-intervals. However, such a physical phenomenon is not possible since the second pulse cannot cause any further breakdown as long as the high-temperature plasma of the first pulse is available at the breakdown location. In contrast to this, the incoming photons of the second pulse can only be absorbed by an inverse-bremsstrahlung process. The breakdown can only be caused by a second pulse when the plasma has cooled significantly and a sufficient number of atoms are ready to ionize. It is, therefore, clear from the above discussion that there is an incomplete understanding of the absorption of the second pulse. To understand the nature of the energy absorption from the second pulse, we performed the DP-ELD experiment in quiescent air. As a result, three different characteristics of the second pulse's energy absorption are found if it is triggered at different pulse intervals.
\begin{enumerate}
	\setcounter{enumi}{0}
	\item It undergoes IB-absorption.
	\item It causes a multi-mode breakdown.
	\item It creates a breakdown similar to the SP-LIB.
\end{enumerate}

These different characteristics are due to the nature of the interaction of second pulse photons with electrons and heavy particles.
Therefore, to avoid confusion, we have used the very generic term laser energy deposition in this study while discussing successive pulses.

\par
This study also witnesses a fascinating and first insight of some new physical phenomena during successive LEDs, i.e. the generation of the fourth-lobe (backward lobe) and multiple shock-waves in DP-LED when the pulse interval is $\geq$ 30 $\mu{s}$ and $\leq$ 3 $ms$. The phenomena of shock-wave propagation, evolution of the plasma kernel (hot plume), and third-lobe (forward lobe) have been described in several studies for SP-LIB \cite{gebel2015optical, dumitrache2017control, nakaya2017flame, phuoc2006laser}. However, the actual reason for the generation of the third lobe is not yet clearly understood, and it is still a debatable point among the researchers \cite{Singh_2019,phdthesisawanish}. 
Therefore, a new approach is used in this study based on Maxwell’s theory of light-matter momentum exchange to explain the previously unexplained (generation of third-lobe) and newly observed (generation of fourth-lobe and multiple shock-waves) physical phenomena. This new approach offers the possibility to calculate radiation pressure for the relativistic and the non-relativistic case and explains the plausible reason for  the generation of both lobes and the propagation of multiple shock-waves.
This study also provides the important additional data of photon-flux and electric-field associated with first and second pulse LED that could be useful for the calculating electron density, breakdown thresholds and IB-coefficients in other future numerical studies. 
The findings of this study may be useful in a variety of scientific and engineering disciplines where control over plasma energy and lifetime are paramount. Some of these are briefly discussed in the following sections.
\par
It is known that weak ionization of gases leads to significant changes in drag reduction and the shock stand-off distance in front of a blunt body in ballistic tunnels. These interesting plasma effects are of great practical importance for high-speed aerodynamics and flight control \cite{joarder2017two, Fridman_2008}. 
Various techniques have recently been used to generate plasma, but unfortunately they have not been used in practice due to the complexity associated with the phenomena and their implementation. The complexity lies in depositing energy in very short intervals (in order of nanoseconds) and controlling it precisely with space and time. Laser-induced plasma (LIP), however, offers spatio-temporal precision and control over energy deposition. Implementing the LIP system is also straightforward since the laser beam can be guided via optical fiber to create the plasma. In metallurgical process, a high temperature is often required and hitherto conventional techniques are used to generate the plasma. Producing plasma by the conventional method requires more space and time to complete the process, and therefore the process becomes inefficient \cite{Fridman_2008}. However, such a plasma can easily be generated via a focused laser beam guided by the optical fiber. The advantage of laser-induced plasma is that it can complete the process in one-stage or directly, which is especially important for reducing metals from their oxides. However, conventional technology requires multiple steps for the same task. LIP can also offer benefits in other processes such as surface cleaning, welding, cutting, printing, engraving, etc.

Plasma has also been used extensively in biology and medicine due to its potential benefits \cite{Vogel_2003, Vogel_2005, Cherian_2008}. Recently, it has been used for surface sterilization and treatment of various surfaces to control their compatibility with bio-materials \cite{Fridman_2008}. Nowadays, it has been used directly in medical applications to treat living tissues and cells, including sterilization and healing of wounds, blood clotting, hair removal, skin diseases and cancer-related tumours. In all of these applications, controlling energy and plasma lifetime is paramount. In this context, the technique for controlling these parameters is presented in this study.

In combustion, the spark plug is one of the oldest techniques for generating plasma and is still used for the same, although it has many disadvantages \cite{phdthesisawanish}. Laser ignition is one of the most effective alternatives to conventional spark plugs, but has not been used in practice due to its incomplete understanding. In recent times, Several studies have reported the advantages of SP-LIB for initiating ignition and combustion \cite{gebel2015optical, dumitrache2017control, nakaya2017flame, phuoc2006laser, Singh_2019, Bradley_2004}. Singh et al. \cite{Singh_2019} presented the associated challenges and solutions for use in practical application. They showed that the lifetime of the plasma kernel governs the occurrence of ignition instead of minimum ignition energy (MIE) or the breakdown spark. They reported that minimum energy is required to create a breakdown spark, but not self-sufficient to trigger ignition. 
In order to improve the lifetime of the plasma kernel, it is required to increase energy significantly with SPLIB. However, increasing pulse energy has some disadvantages. First, it overkills the breakdown zone \cite{Bradley_2004}, and second, it cannot be delivered with low damage threshold optical devices (e.g. fiber optic cables, optical lens). \cite{phdwermer}. Therefore, in practice, the desired liftime of the plasma kernel cannot be achieved with SP-LIB due to the problems mentioned above. To avoid such a hurdle, we must either improve the damage threshold of the optical devices or use a new method that works with the existing optical systems. Improving damage threshold require further research on optical materials. However, using the new technique is more reliable and efficient. In this regard, DP-LED is used in this study, which offers the advantage of delivering high energy through successive pulses with existing optical fiber. This study aims to present the pulse interval scheme between the successive pulses with which the DP-LED is effective for practical use.


\par
In the applications mentioned earlier, a clear understanding of plasma core behaviour is required to determine its reaction rate with the surrounding medium (solid, liquid, or gas) in different successive LEDs. Therefore, in this regard, three plasma kernel parameters (width, height, and projection area) are measured with the in-house MATLAB algorithm from schlieren images. These parameters are also calculated in other studies to calculate the mixing and expansion rate of the plasma kernel \cite{bak2015successive, an2017experimental}.
The measured parameters are further used to calculate the expansion speed of the plasma kernel along the height and width. However, from the measured parameters (width, height and projection area) in the present study and the other previous studies \cite{bak2015successive, an2017experimental}, it is not clear that which one of SP-LIBs and DP-LEDs are expanding faster. 
Therefore, to determine the actual expansion of the plasma, the volume and volumetric expansion rate (VER) are calculated, which ultimately help estimate the time plasma-kernel would take to trigger the phenomenon at various combinations of energy and pulse interval.

\section{Materials and Methods}
\label{S:2}
\begin{figure}[bt!]
\centering
\includegraphics[trim={0bp 0bp 0bp 0bp},clip, width=0.85\columnwidth]{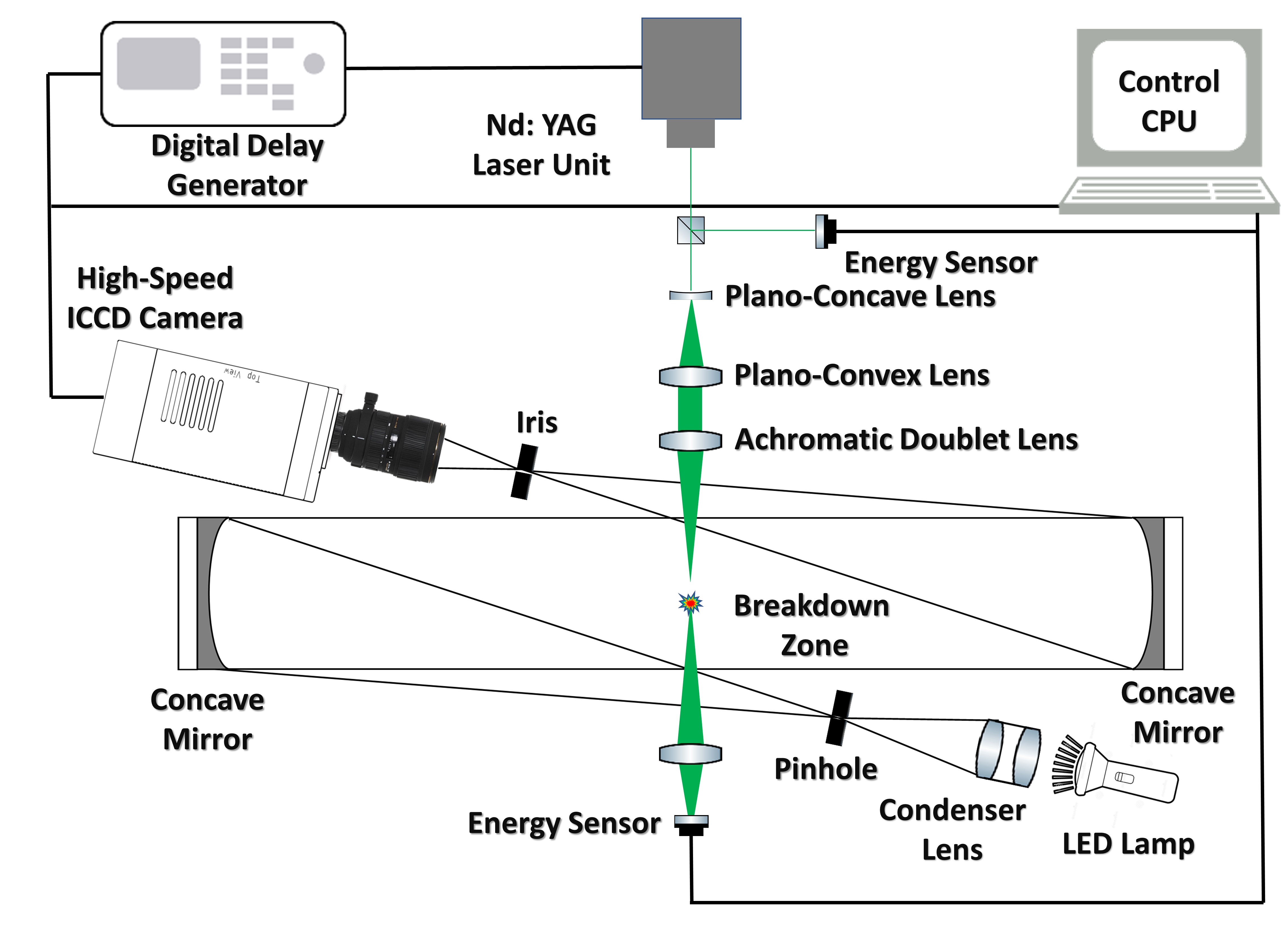} 
\caption{\label{fig:Fig0}\small Schematic of the experimental setup to visualize the propagation of the shock wave and the evolution of the plasma kernel after the laser-induced breakdown.}
\end{figure}

LIB plasma in quiescent atmospheric air was produced by Q-switched Nd: YAG laser (Quantel EVG00200) having dual cavities. The laser was capable of producing two pulses at a very short intervals (in nanoseconds) in the range of 10-200 mJ per pulse at a wavelength of 532 $nm$ with a pulse duration of 7 $ns$. The pulse to pulse energy variation of the laser was less than 3\%. The laser pulse was steered with a laser arm, and then it was expanded and collimated to a diameter of approximately 30 mm by a Galilean beam expander in order to produce a tighter focus. To precisely capture the LIB phenomena, a reflect-type high-speed schlieren imaging system was used. Schlieren images were recorded with a high-speed ICCD camera (4 Quick E, Stanford computer optics). The camera was operated at an exposure time of 20 $ns$. The shorter gating time allowed the precise recording of the physical phenomena. As a result, the repeatability of the experiment was very good. The spatial resolution (78.125 $\mu{m}/pixel$) of the image recorded during the experiment was determined by using a calibration target. A digital delay generator (BNC-745) was used for triggering the camera, laser system, and the other electronic devices. The energy during the experiment was measured by using a pyroelectric energy sensor (Thorlabs, ES120C). The arrangement of the experimental setup is shown schematically in 
Fig.~\ref{fig:Fig0}.

\section{Results and Discussion}
\subsection{Characteristics of successive laser energy deposition}
\begin{figure}[bt!]
\centering
\includegraphics[trim={20bp 10bp 252bp 71bp},clip, width=0.45\columnwidth]{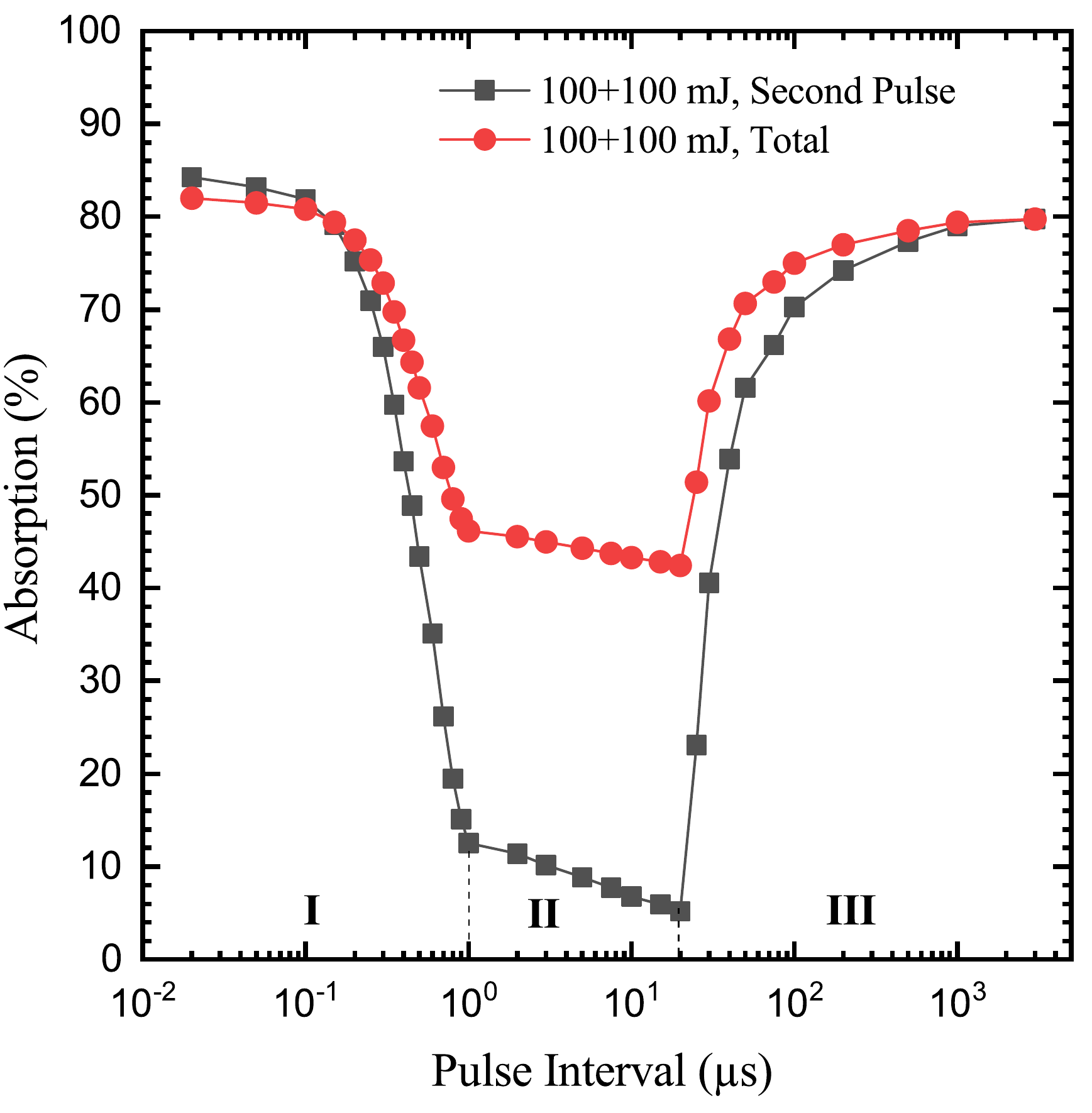} 
\caption{\label{fig:Fig1}\small Percentage of absorbed energy from only second-pulse and total (both pulses together) in 100+100 mJ DP-LED at different pulse-intervals ranging between 20 $ns$ to 3 $ms$.}
\end{figure}

Figure~\ref{fig:Fig1}  shows the combined energy absorption from both pulses and only from the second pulse measured during the DP-LED. In the case of DP-LED, the energy absorption of the second pulse is measured as the difference between the combined absorption and that of the first pulse. The percentage absorption shown in Fig.~\ref{fig:Fig1} is obtained by averaging three different breakdown for each pulse interval. The reproducibility is excellent because the measured variation from pulse to pulse energy is less than three percent. Therefore, no error bars are shown in Fig.~\ref{fig:Fig1}. 

The incident energy of 100+100 mJ is used at various pulse-intervals between 20 $ns$ to 3 $ms$. It is observed that DP-LED is only effective up to a pulse-interval of $\Delta{t_{p}}$ = 150 $ns$ with higher absorption (79.89 \%) compared to SP-LIB (79.75 \%). Between 150 $ns$ to 1 $\mu{s}$ pulse-interval, the absorption of second-pulse decreases drastically and reaches 12.55 \% at 1 $\mu{s}$. After a pulse-interval of 1 $\mu{s}$, the absorption continues to decrease gradually, reaching the lowest 5.18 \% at 20 $\mu{s}$. After 20 $\mu{s}$, the absorption quickly increases up to 200 $\mu{s}$ (74.19 \%) and later gradually up to 3 $ms$. After 3 $ms$, the breakdown created by the second-pulse is similar to the first-pulse breakdown, since the plasma kernel created by the first breakdown is cooled by the surrounding medium and almost reaches the initial state. The absorption of the second pulse in DP-LED is superior to SP-LIB for the pulse interval of less than 150 $ns$, since during this interval the energy absorption is dominated by electron-ion inverse-bremsstrahlung (IB) over the electron-neutral IB process. With the same incident energy, however, the total absorption of SP-LIB (87.6 \%, 200 mJ Single) is higher than the maximum of DP-LED ( 84.23 \%, 100+100 mJ at $\Delta{t_{p}}$ = 20 $ns$). This is because in the case of SP-LIB, the plasma does not have time to expand and take away the ions, atoms and molecules. Based on the absorption characteristics at different pulse-intervals, this study has categorized the absorption from the second pulse into three zones.  
The range of pulse-intervals for zone-I (20 $ns$ to 1 $\mu{s}$), zone-II (1 $\mu{s}$ to 20 $\mu{s}$), and zone-III (20 $\mu{s}$ to 3 $ms$) are shown in Fig.~\ref{fig:Fig1}. 
In zone-I and zone-II, the second-pulse does not create a breakdown, it only absorbs energy through IB-absorption process by the high temperature plasma. However, it has been found that the second breakdown (multi-mode) occurs in zone-III because the plasma in this zone has cooled rapidly.

The percentage of absorption in successive LEDs can be assessed by the following relationship:
\begin{equation}
\left(\alpha_{t}\right)_{n} = \sum_{i=1}^{n}\frac{E_{i}}{(E_{p})_{n}}\left(\alpha_{i}\right)_{\Delta t_{p}(i-1\rightarrow i)}
\label{eq:three}
\end{equation}
where, $(\alpha_{t})_{n}$ and $(E_{p})_{n} = {\sum^{n}_{i=1} E_{i}}$ are the total absorption (\%) and incident pulse-energy (mJ) of the n-pulses, $(\alpha_{i})_{\Delta t_{p}(i-1\rightarrow i)}$ is the percentage absorption of $i^{th}$ pulse corresponding to energy $E_{i}$ for the pulse-interval ${\Delta t_{p}(i-1\rightarrow i)}$.
For successive laser energy deposition of two pulses, i.e. for $n=2$,
on right side of the Eq.~(\ref{eq:three}), $i=1$, $(\alpha_{i})_{\Delta t_{p}(i-1\rightarrow i)} = (\alpha_{1})_{\Delta t_{p}(0\rightarrow1)}$ represents the percentage absorption from first-pulse, which corresponds to the absorption during the SP-LIB ($ \alpha_{1}$) and $i=2$, $(\alpha_{i})_{\Delta t_{p}(i-1\rightarrow i)} = (\alpha_{2})_{\Delta t_{p}(1\rightarrow2)}$ represents the percentage absorption from second-pulse. The absorption of each pulse can also be determined up to $i=n$. Therefore, the total energy absorbed in successive n-pulse LEDs can be determined by $(E_{a})_{n} = (E_{p})_{n}\,({\alpha_{t}})_{n}$. 

\begin{table}[t!]
	\centering
	\caption{\label{tab:table2}\small Absorbed energy ($E_{a}$) and the percentage absorption ($\alpha$) out of incident pulse-energy ($E_{p}$)  and the calculated shock-wave energies, $E_{s}$ (by Jones model \cite{gebel2015laser}) and radiation pressure (${P_{rad}}$) (by Eq.~(\ref{subeq:2})) for SP-LIBs and DP-LEDs.
	}
	\begin{tabular}{cccc}
		\hline
		$E_{p}$ (mJ),&$E_{a}$ (mJ),&\mbox{$E_{s}$ (mJ)},&\mbox{${P_{rad}} \times 10^{3}$}\\
		$\Delta{t_{p}} \:(\mu{s})$&$\alpha \: (\%)$&\mbox{shock-loss (\%)}&\mbox{(N/m$^{2}$)}\\
		\hline
		
		100, Single&79.75, 79.75&\mbox{63.50, 79.62}&\mbox{396.28}\\
		200, Single&175.2, 87.6&\mbox{142.75, 81.57}&\mbox{870.58}\\
		100+100, 0.05&162.93, 81.47&\mbox{142.57, 87.50}&\mbox{396.28+413.33}\\
		100+100, 0.1&\mbox{161.63, 80.81}&\mbox{139.43, 86.27}&\mbox{396.28+406.86}\\
		100+100, 50&141.31, 70.66&\mbox{120.14, 85.02}&\mbox{396.28+305.91}\\
		100+100, 100&\mbox{150.03, 75.02}&\mbox{128.16, 85.42}&\mbox{396.28+349.23}\\
		\hline
	\end{tabular}
\end{table}

\begin{figure}[bt!]
	\centering
	\includegraphics[trim={0bp 0bp 0bp 0bp},clip, width=1\columnwidth]{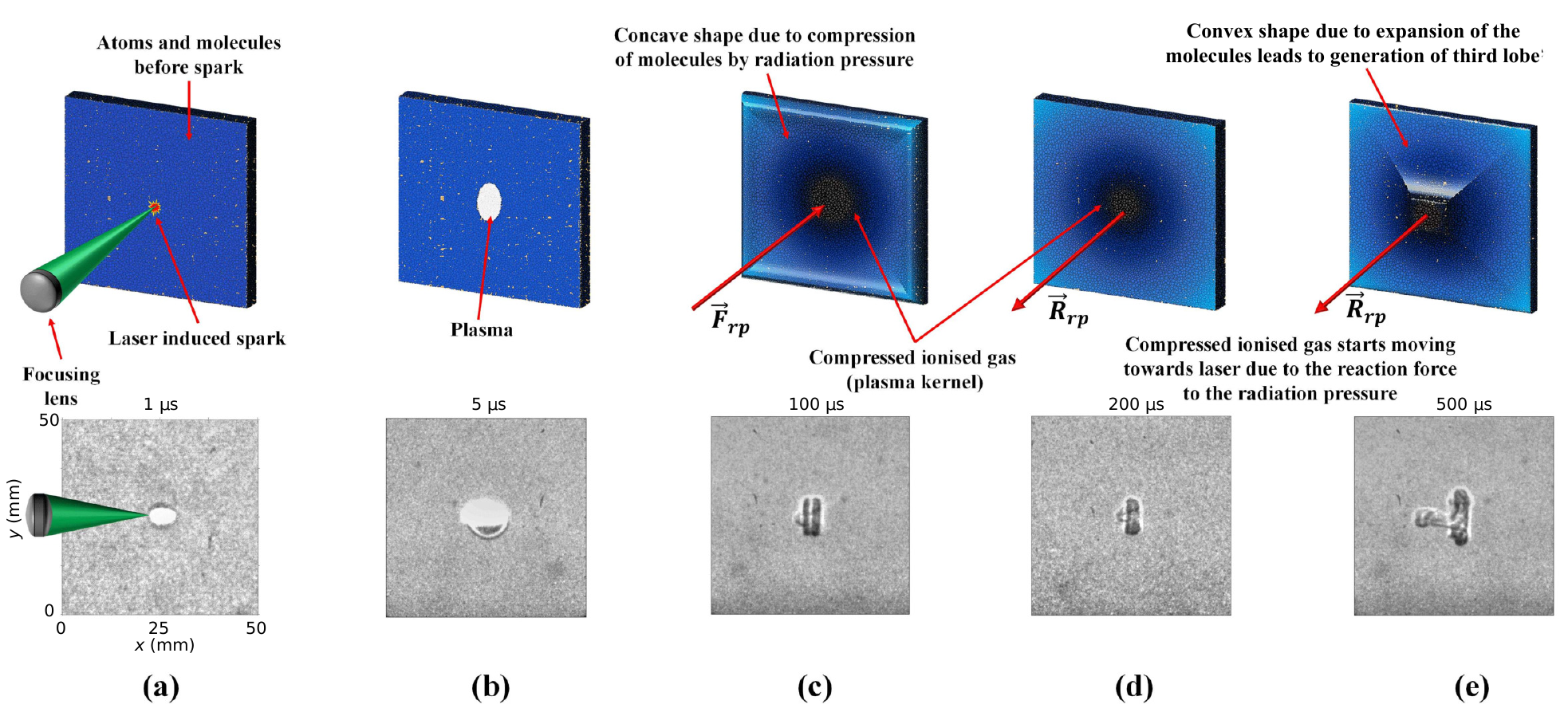}
	\caption{\label{fig:sh1}\small Schematic and experimental images describing the mechanism of energy absorption and its subsequent effect on the development of the plasma kernel. (a) energy deposition through MPI and subsequent absorption via inverse bremsstrahlung process, (b) further growth of plasma due to cascade ionization, (c) radiation pressure due to plasma shielding effect and formation of compressed ionized gas, subsequently, (d) the compressed gas begins to retract due to reaction force to the radiation pressure, (e) the reaction force leads to the development of the lobe (third-lobe) opposite to the direction of the laser.}
\end{figure}

\begin{figure}[bt!]
	\centering
	\includegraphics[trim={0bp 0bp 0bp 0bp},clip, width=1\columnwidth]{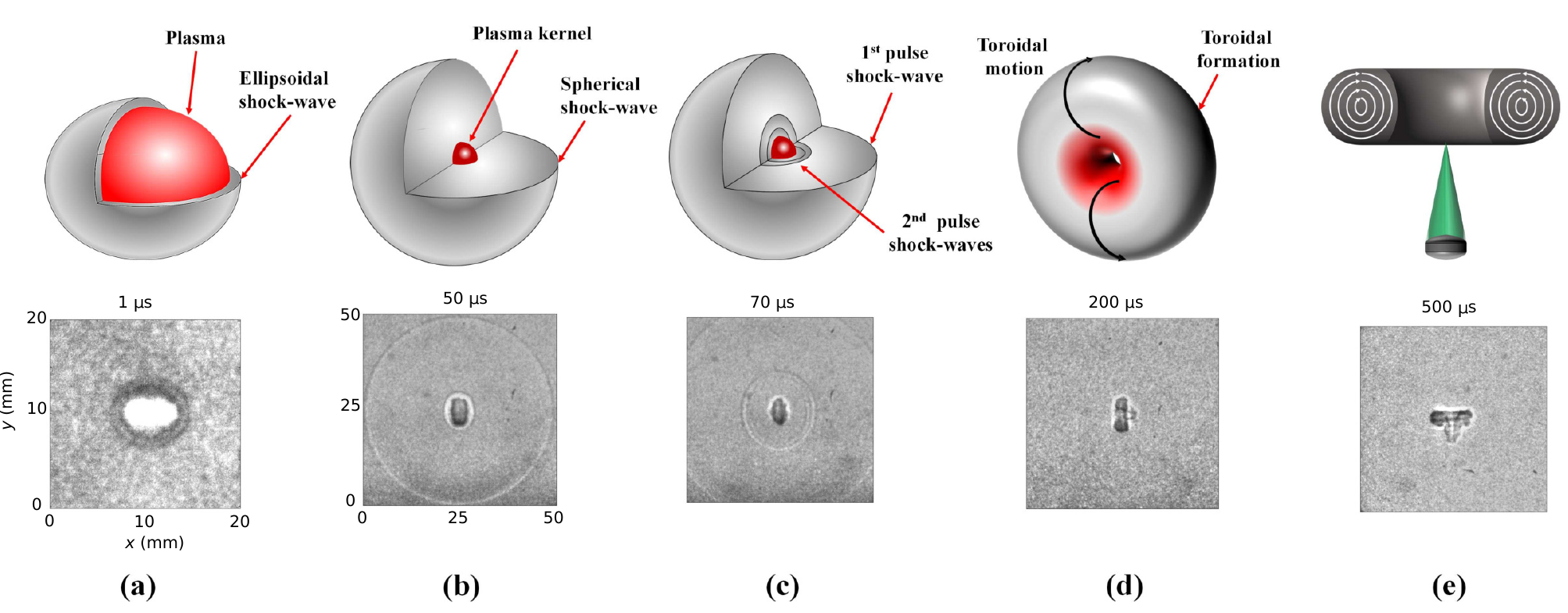}
	\caption{\label{fig:sh2}\small Schematic and experimental images describing the shock evolution and the formation of the toroidal-ring after a laser-induced breakdown. (a) propagation of ellipsoidal shock together with high-temperature plasma $\sim 1-3 \; \mu s$, (b) After separation of the shock wave from the plasma, the shock propagates radially outward as a spherical shock and the plasma cools to form compressed ionized gas, (c) The multi-mode breakdown occurs when the second pulse is triggered after $\sim 30 \; \mu s$ µs due to the presence of compressed ionized gas, which significantly changes the breakdown threshold, and therefore the multiple shock waves begin to propagate from that location, (d-e) formation of the toroidal ring. } 
\end{figure}

The energy absorption from the second pulse changes in the DP-LED when the pulse interval changes after the first-pulse LIB. This is because the first-pulse LIB generates a perturbation in the density field in such a way that the incoming second laser-pulse experiences a different optical density at the breakdown location  (when temporally separated between 20 $ns$ and 3 $ms$), as shown in Fig.~\ref{fig:Fig1}.
The absorption from the second pulse therefore also describes the optical transparency that is generated by the first breakdown at different temporal positions. The optical transparency increases in zone-I, since after the multi-photon ionization by the first pulse, the seed electrons begin to absorb incoming photons through IB-absorption. The IB-absorption then increases the kinetic energy of the electrons. The high-energy electrons further ionize other nearby atoms and molecules and ultimately lead to an electron avalanche. This overall process creates the first breakdown and triggers a radial outward movement of high-energy plasma and shock wave. The plasma and shock wave sweeps the electrons, ions, atoms, and molecules away from the breakdown location. As a result, the optical density at this point decreases over time, which eventually leads to a decreased absorption during the second pulse LED. The absorption decreases with increasing pulse interval until the plasma and shock wave move together, i.e. $\sim$ 1 $\mu{s}$, after which the separation occurs. Due to separation, the energy left with plasma is very small, as the shock wave takes away approximately 79.62 \% of the total absorbed energy of the first-breakdown (Table~\ref{tab:table2}). For a better understanding, a schemeatic of the entire event are shown in Fig.~\ref{fig:sh1} and~\ref{fig:sh2}.

Zone II begins with the separation of plasma and shockwave. In this zone, the optical transparency increases (i.e. breakdown threshold for second pulse increases) gradually between 1 $\mu{s}$ and 20 $\mu{s}$ due to outward movement of the plasma leading to its lower density at the core. But even after detachment with the shock wave, the plasma moves radially outwards up to 20 $ \ mu {s} $ due to the initial momentum. After that, the plasma begins to cool and retreat inward due to the lower core density and pressure.
 Zone III begins after 20 $\mu{s}$, when the plasma begins to cool down rapidly. It consists of partially compressed ionized gas. As a result, the optical transparency decreases, and the absorption of the second pulse increases. The probable reason for the above physical phenomenon can be understood by the Saha ionization equation, 
\begin{equation}
\frac{n_{e}n_{i}}{n_{a}}= C^{'} T^{{3}/2}\,e^{{-W}/{k_{b}T}}
\label{eq:four}
\end{equation}
where $n_{e}$, $n_{i}$, and $n_{a}$ are the number density of electrons, ions, and atoms respectively. $W$ and $K_{b}$ are the potential energy and Boltzmann constant, and $T$ is the temperature of plasma in thermal equilibrium. $C^{'}$ is the constant that depends on the atomic cross-section, plank constant and other constant factors \cite{bataller2014blackbody, Joarder_2021, Alberti_2019}.
Let us now understand what is really happening physically and how  Eq.~(\ref{eq:four}) can explain the physical phenomena that cause the change in optical transparency in all three zones after the first-breakdown. In the above equation, the product $n_{e}n_{i}$ represents the rate of recombination, and the rate of ionization depends on the number of atoms $n_{a}$ that can be ionized. In the initial state, when there is no energy deposition, the process is neutral, and therefore these two rates are the same. Now let's see what happens when the first pulse is triggered.

\begin{figure}[bt!]
	\centering
	\includegraphics[trim={0cm 0cm 0cm 0cm},clip, width=0.95\textwidth]{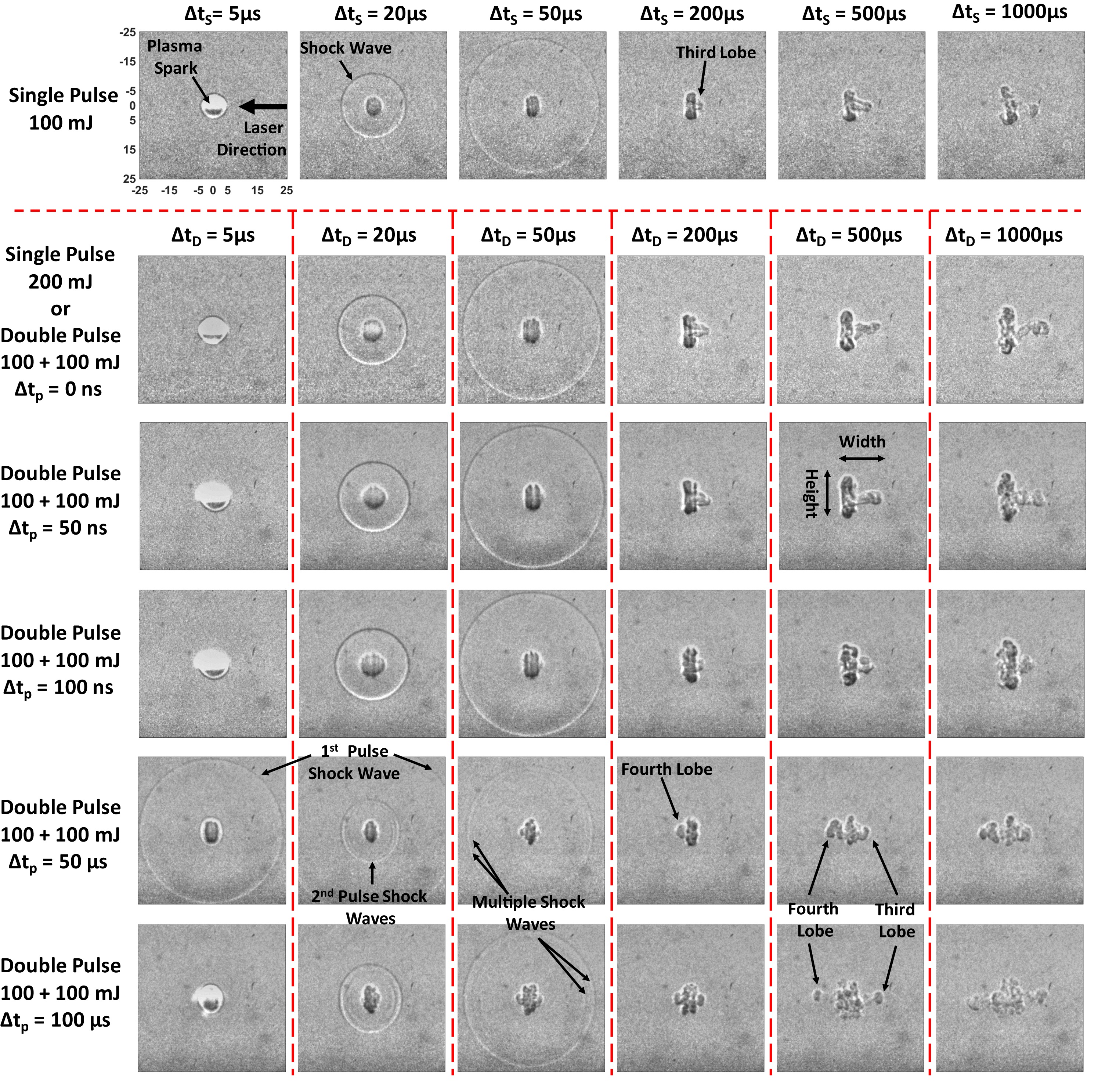}
	\caption{\label{fig:Two}\small Schlieren images of SP-LIBs (100 mJ and 200 mJ) and DP-LEDs ( 100+100 mJ and $\Delta{t_{p}}$ = 0 $ns$, 50 $ns$, 100 $ns$, 50 $\mu{s}$, 100 $\mu{s}$) in the quiescent air. The DP-LED with $\Delta{t_{p}}$ = 0 can be also considered as SP-LIB, if both the laser pulses are same. The direction of laser is shown in the image (100 mJ, $\Delta{t_{S}}$ =  5 $\mu{s}$), where (0, 0) is the focal point of the laser. The actual size of each image is 50 $mm$ $\times$ 50 $mm$. $\Delta{t_{S}}$ = $\Delta{t_{p}}$ + $\Delta{t_{D}}$, where, $\Delta{t_{p}}$ is pulse-interval, and  $\Delta{t_{S}}$ and $\Delta{t_{D}}$ are the time after first and second pulse LED.}
\end{figure}

Zone-1 begins immediately after the first pulse is fired, and then the electrons receive energy from the incident photons due to their low mass and high mobility. They then collide and transfer the energy to all other heavy particles for the ionization and maintain $n_{e}n_{i}/n_{a} \ll 1$. As a result, the local temperature and pressure of the plasma increases and begins to move radially outward. Due to outward movement of the plasma, the density of electrons and heavy-particles decreases at the breakdown location. In later times, the temperature of the plasma further increases due to the higher likelihood of electron-ion interaction leading to recombination, but the optical transparency continues to decrease due to lower available density at the core.
A few hundred nanoseconds after energy deposition, when electrons and ions encounter each other, recombination occurs, and in some cases, ionization. The rate of recombination and ionization depends on the energy state of the electron during the interaction. The rate of such processes can be described with the help of the electron energy distribution function (EEDF) $f(e)$, which is the probability density of electrons whose energy is $e$ \cite{Fridman_2008}.
Zone-II starts with the separation of plasma and shock wave, and the shock wave takes away most of the absorbed energy. However, the process has almost reached neutral  $n_{e}n_{i}/n_{a}  \simeq 1$ but the plasma still moves outward up to $\sim$20 $\mu{s}$ due to the initial momentum caused by first pulse LIB. Later, the plasma begins to retreat due to its low available energy and also due to the lower pressure in the core. The fluctutation of core pressure can also be seen in Fig 16 also in our previous works \cite{Joarder_2021, Joarder_2013}.  
Zone III starts when the plasma begins to cool and recombination begins to dominate the ionization process $n_{e}n_{i}/n_{a} \geq 1$.  After 20 $\mu{s}$, the temperature begins to drop rapidly, and plasma moves toward the core to form a compressed ionized gas (Fig.~\ref{fig:Two}). Therefore, The increased rate of recombination then increases the atomic density on the core, which ultimately leads to a lower optical transparency. As a result the absorption from the second pulse increases rapidly. The increase in absorption leads to a multimode breakdown when a second pulse is triggered during this period. The multimode breakdown continues until the process reaches its initial state (similar to the first pulse energy deposition or the atmospheric conditions).

\subsection{Development and Evolution of Plasma Kernel}

\begin{figure}[bt!]
 	\centering
 	\includegraphics[trim={0bp 0bp 0bp 0bp},clip, width=1\columnwidth]{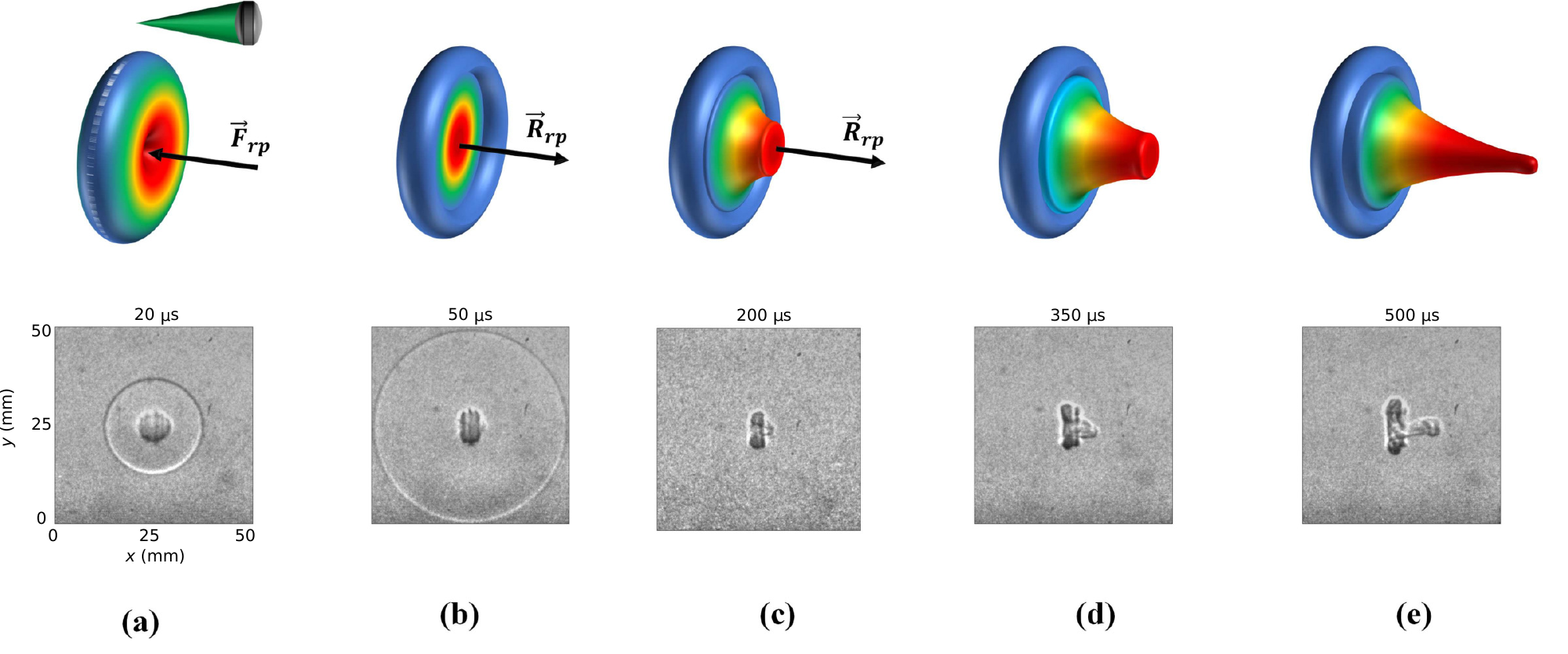}
 	\caption{\label{fig:sh3}\small  Description of plasma kernel evolution after a single-pulse laser-induced breakdown. Here, $\overrightarrow{F}_{rp}$ is the compressive force caused by the radiation pressure in the breakdown zone due to the plasma-shielding effect, and  $\overrightarrow{R}_{rp}$ is the reaction force to $\overrightarrow{F}_{rp}$. A comparison of the experiment with the numerical simulation for the single-pulse laser induced breakdown is shown in Fig.~\ref{fig:single_pulse}.}
 \end{figure}

Fig.~\ref{fig:Two} shows the experimental images of SP-LIBs and DP-LEDs in quiescent air. SP-LIB creates a strong electric-field that causes intense heating of neighboring free electrons and heavy particles (ions, atoms, molecules, etc.), and leads to a rapid hydrodynamic expansion of the plasma and the shock wave (see Fig.~\ref{fig:Two}, 100 mJ and 200 mJ Single). Nearly after 1 $\mu{s}$, the plasma and the shock wave separate from each other, and the shock wave takes away most of the absorbed energy.  However, the plasma is still expanding due to the left over energy and reaches a constant size at $\sim$ 20 $\mu{s}$. After 20 us, it begins to retreat along the width due to the radiation pressure \cite{Astrath_2014}, and this phenomenon continues till $\sim$ 100 $\mu{s}$.  Then a third lobe (forward lobe) is generated against the direction of the laser beam (see Fig.~\ref{fig:Two}, $\Delta{t_{S}}$ = 200 $\mu{s}$, 100 mJ Single), which causes rapid and unstable growth of the plasma kernel. A complete description of SP-LIB is shown in Fig.~\ref{fig:sh3}.
 Several studies have reported the generation of third-lobe in SP-LIBs, but the actual cause of the generation of the lobe has not yet been understood and it is still a debatable point among researchers.
 \par

\begin{figure}[bt!]
	\centering
	\includegraphics[trim={0bp 0bp 0bp 0bp},clip, width=1\columnwidth]{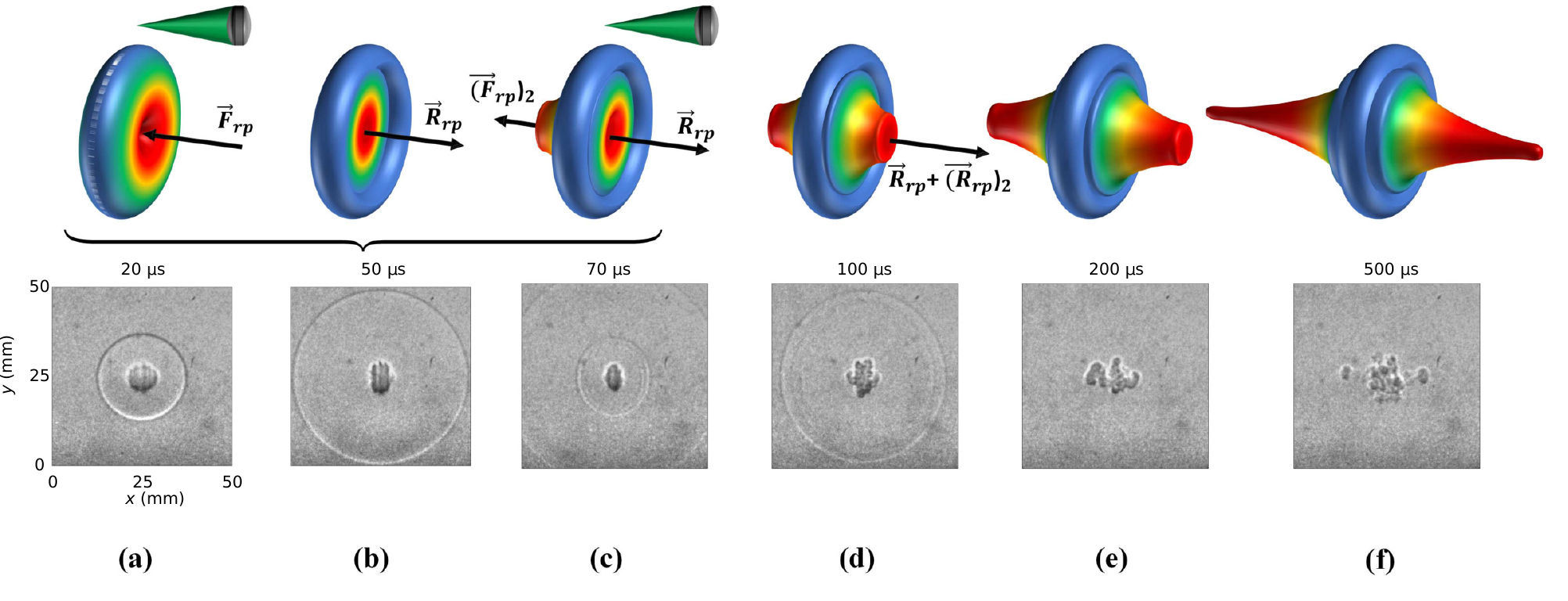}
	\caption{\label{fig:sh4}\small Description of the development of the plasma kernel during the successive laser energy deposition ( at $\Delta{t_{p}}$ =  50 $\mu{s}$). A comparison of the experiment with the numerical simulation for the successive laser energy deposition is shown in Fig.~\ref{fig:double_pulse}.}  
\end{figure}

For DP-LEDs with incident energies of 100+100 mJ for $\Delta{t_{p}}$ = 50 $ns$ and 100 $ns$,  the observed physical phenomena are similar to those for SP-LIBs.  While for $\Delta{t_{p}}$ = 50 $\mu{s}$ and 100 $\mu{s}$, the physical phenomena are not similar to SP-LIB, and some new and interesting phenomena are observed, which is a first direct observation of the generation of the fourth-lobe (backward lobe) and multiple shock-waves (see Figures ~\ref{fig:Two} and ~\ref{fig:sh4}).
To understand the actual reason why both lobes and multiple shock waves are generated, this study uses a new approach based on Maxwell's prediction of light-matter momentum exchange. This approach suggests that the third lobe in SP-LIBs is generated against the direction of the laser beam due to the reaction force ($R_{rp}$) to the radiation pressure induced by laser sparks at the focal point (see Fig.~\ref{fig:sh1}), which is analogous to a drop falling into a pool \cite{Shankar_1995,Cai_1989}. The multiple shock waves propagating outwards are generated in DP-LEDs because the radiation pressure induced by the second pulse LIB interacts at multiple points within the perturbed density field of the plasma kernel (see Fig.~\ref{fig:sh2} and~\ref{fig:Two}). The fourth lobe is generated in the laser beam direction due to the force exerted by radiation $(F_{rp})$ on optically thick compressed plasma kernel. The fourth lobe and multiple shock waves are observed in this study when the second pulse is triggered between the 30 $\mu{s}$ to 3 ${ms}$.

After a pulse interval of about 50 $\mu{s}$, the fourth lobe first appears in the direction of the laser beam (see Fig.~\ref{fig:Two}, $\Delta{t_{D}}$ = 200 $\mu{s}$, for $\Delta{t_{p}}$ = 50 $\mu{s}$ ), and then third lobe appears against the direction of laser beam (see Fig.~\ref{fig:Two}, $\Delta{t_{D}}$ = 500 $\mu{s}$, for $\Delta{t_{p}}$ = 50 $\mu{s}$ ). The third and fourth lobes appear simultaneously when the pulse interval is $\sim$ 100 $\mu{s}$ (see Fig.~\ref{fig:Two}, $\Delta{t_{D}}$ = 200 $\mu{s}$, for $\Delta{t_{p}}$ = 100 $\mu{s}$). This is because the third lobe is ready to appear before the second pulse is deposited. It is also observed that the generation of the third lobe is delayed by the second pulse LIB when $\Delta{t_{p}} <$ 100 $\mu{s}$. However, for $\Delta{t_{p}} >$  100 $\mu{s}$, the second pulse LIB cannot delay the generation of the third lobe, since it has already appeared at $\Delta{t_{S}}$ = 100 $\mu{s}$. For a better understanding, the quantitative analysis and discussion of radiation pressure is presented in the following section.

\begin{figure}[bt!]
\centering
\includegraphics[trim={20bp 10bp 100bp 71bp},clip, width=0.60\columnwidth]{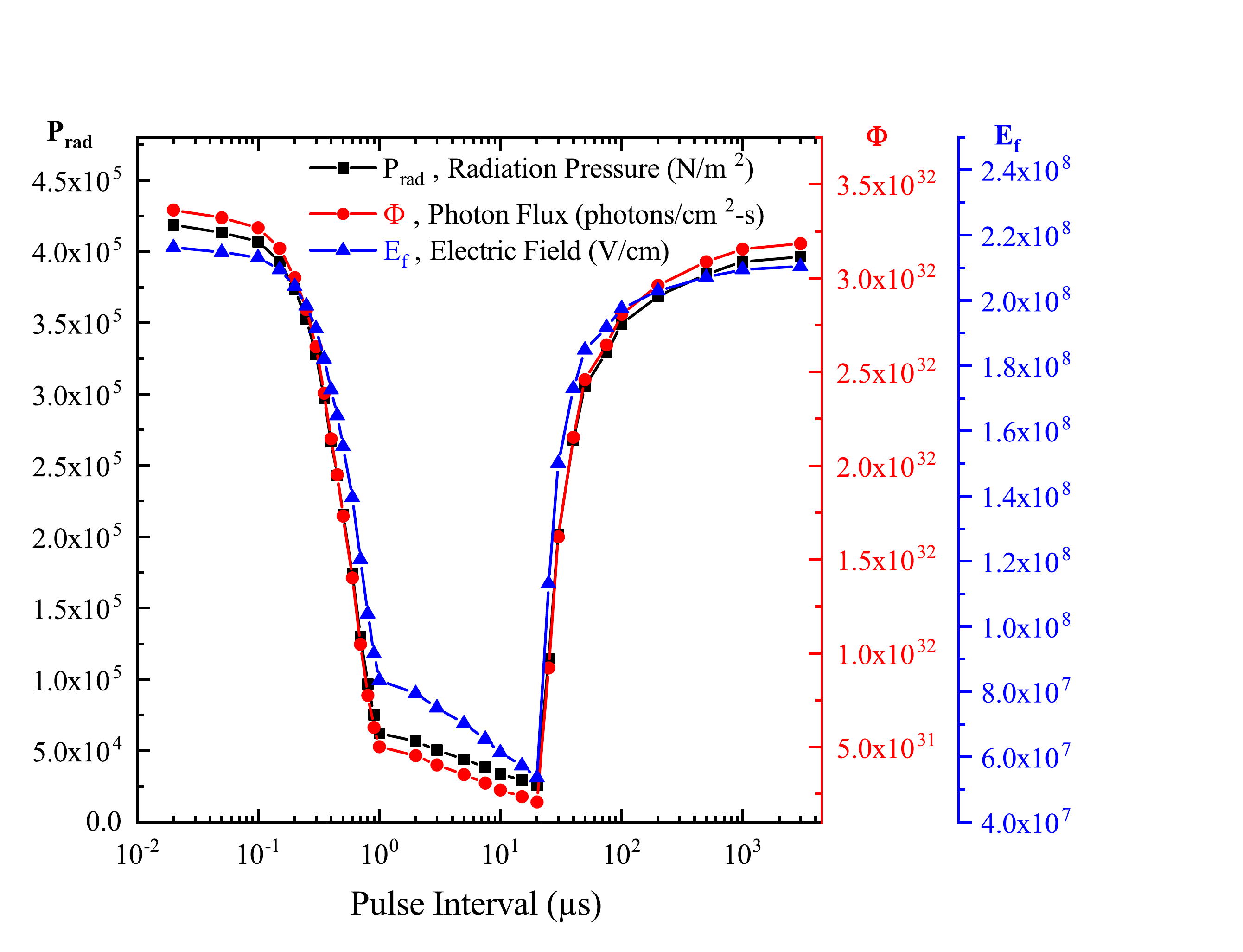} 
\caption{\label{fig:Three}\small Radiation pressure (${P_{rad}}$), Photon-Flux ($\Phi$), and Electric-Field (${E_{f}}$) of second-pulse LED at different pulse-intervals (20 $ns$ to 3 $ms$).}
\end{figure}

In 1871, Maxwell predicted that electromagnetic waves would carry momentum and exert pressure on interacting surfaces in the direction of wave propagation. This prediction was later confirmed in 1901 by Lebedev \cite{lebedew1901untersuchungen} and in 1903 by Nichols and Hull \cite{nichols1903pressure}. The radiation pressure (${P_{rad}}$) due to LIB can be written as \cite{mansuripur2004radiation},
\begin{equation}
{P_{rad}} = \frac{I_{abs}}{c} + 2 \frac{I_{ref}}{c} 
\label{subeq:2}
\end{equation}
where, ${I_{abs}} = {\alpha^{c}} {I_{inc} }$ and ${I_{ref}} = {\beta^{c}} {I_{inc}} $ are the absorbed and reflected intensities, $\alpha^{c}$ and $\beta^{c}$ are the absorption and reflection coefficient, and $c$ is the speed of light in air.
In the LIB study, it is reported that the reflection is negligible \cite{phuoc2006laser}, and the radiation pressure is therefore only induced by absorption. Recently, It has also been reported that shock waves and plasma propagate relativistically up to a few hundred nanoseconds after LIB \cite{Eliezer_2014, 2019arXiv190111377S}. Therefore, if the second pulse is deposited in a very short interval, it will interact with relativistic particles. The radiation pressure for the relativistic case can be determined by the following relationship:
\begin{equation}
P_{rad} = \frac{1}{3} {{a}{T^{4}}} 
\label{subeq:3}
\end{equation}
where, ${a} = {7.5657 \times 10^{-16}} $ J\,m$^{-3}$\,K$^{-4}$ is the radiation constant and $T$ is the plasma temperature. However, the present study does not consider the relativistic effect and the radiation pressure is calculated using Eq.~(\ref{subeq:2}), which is presented in Table~\ref{tab:table2} and Fig.~\ref{fig:Three}.

In this study, we have also calculated the photon-flux and electric-field for SP-LIBs and DP-LEDs. In SP-LIB, the calculated photon-flux is 3.18 $\times$ 10$^{32}$ photons/cm$^{2}$-s for 100 mJ and 6.99  $\times$ 10$^{32}$ photons/cm$^{2}$-s  for 200 mJ, and the electric-field is 2.10 $\times$ 10$^{8}$ $V/cm$ for 100 mJ and 3.12  $\times$ 10$^{8}$ $V/cm$ for 200 mJ. The photon flux and the electric field due to the absorption of the second pulse are shown in Fig.~\ref{fig:Three}. The photon flux is very useful in determining electron density, breakdown threshold, and various other coefficients in physics and in various engineering applications. The electric field is useful in determining thermal and non-thermal plasma. It is determined by the temperature difference between electrons and heavy neutral particles due to Joule heating in the weakly ionized plasma, which is proportional to the square of the ratio of the electric field ($E_f$) to the pressure ($p$). The temperatures of electrons and heavy particles have to be identical for the local thermodynamic equilibrium (LTE) and it is only possible if the $E/p$ values is very small. The temperature difference between them exists because, the electrons after the LIB receive energy from the electric field on their mean free path and lose only a small part of their energy in the subsequent collision with a heavy particles, since they are much lighter than the heavy particles. Therefore, their temperature in the plasma is initially higher than that of the heavy particles. Later, collisions of electrons with heavy particles (Joule heating) can equalize their temperatures, unless the time or energy is insufficient for equilibrium.
 \par
 \begin{figure*}[ht!]
        \centering
        \begin{subfigure}[b]{0.475\textwidth}
            \centering
            \includegraphics[trim={50bp 20bp 56bp 56bp}, width=\textwidth]{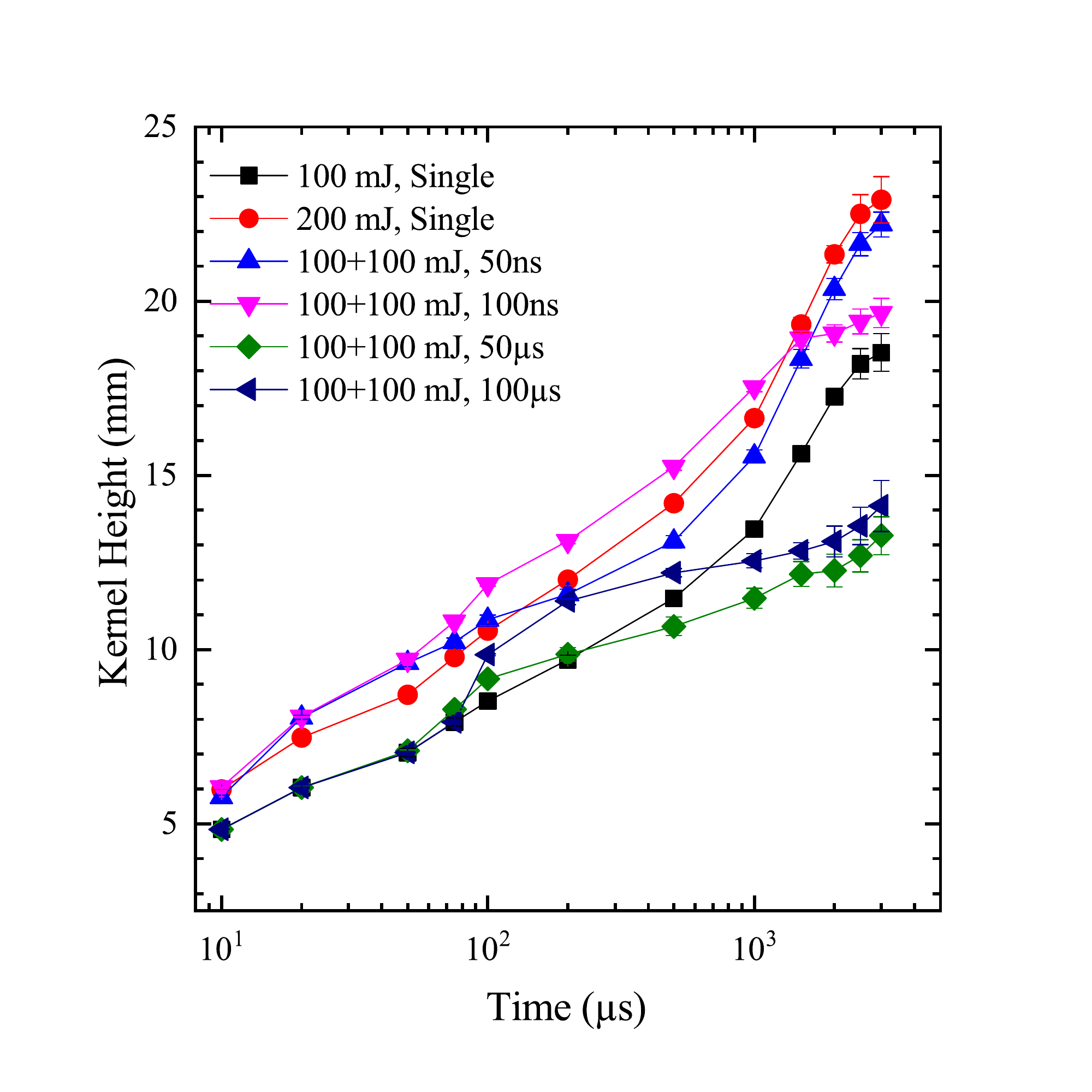}
            \caption[a]%
            {{\small }}    
            \label{fig:height}
        \end{subfigure}
        \hfill
        \begin{subfigure}[b]{0.475\textwidth}  
            \centering 
            \includegraphics[trim={50bp 20bp 56bp 56bp}, width=\textwidth]{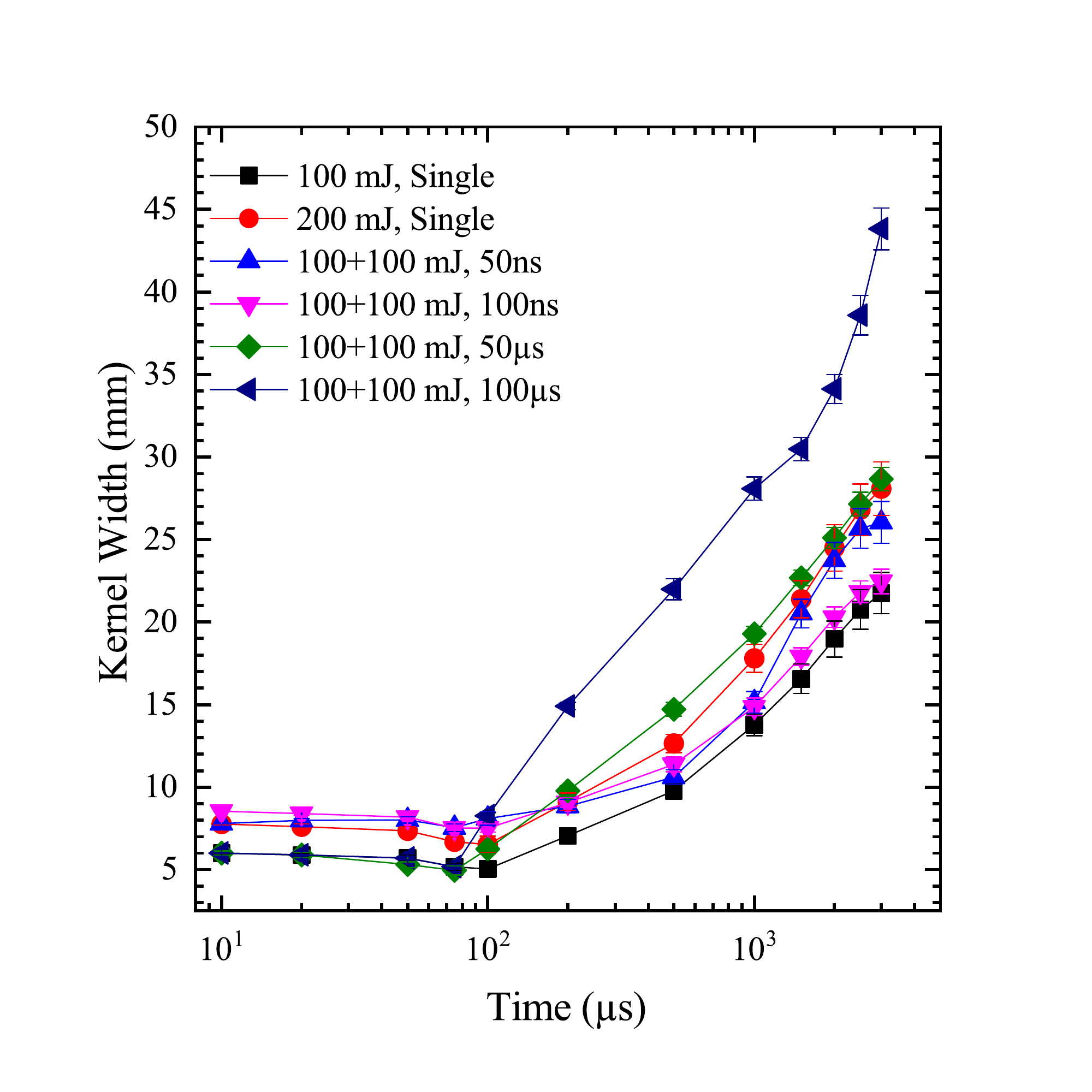}
            \caption[]%
            {{\small }}    
            \label{fig:width}
        \end{subfigure}
        \vskip\baselineskip
        \begin{subfigure}[b]{0.475\textwidth}   
            \centering 
            \includegraphics[trim={50bp 20bp 56bp 56bp}, width=\textwidth]{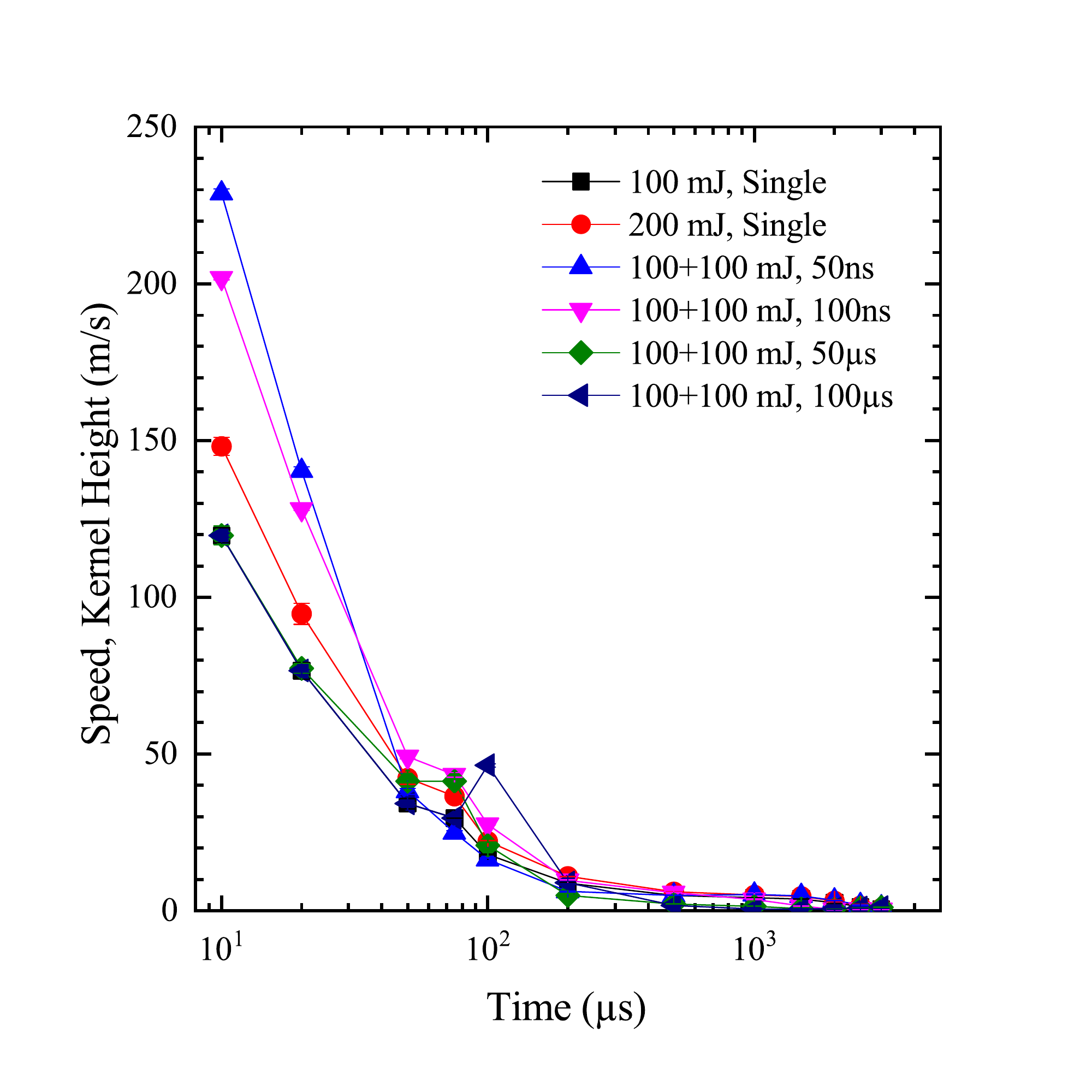}
            \caption[]%
            {{\small }}    
            \label{fig:speedhight}
        \end{subfigure}
        \quad
        \begin{subfigure}[b]{0.475\textwidth}   
            \centering 
            \includegraphics[trim={50bp 20bp 56bp 56bp}, width=\textwidth]{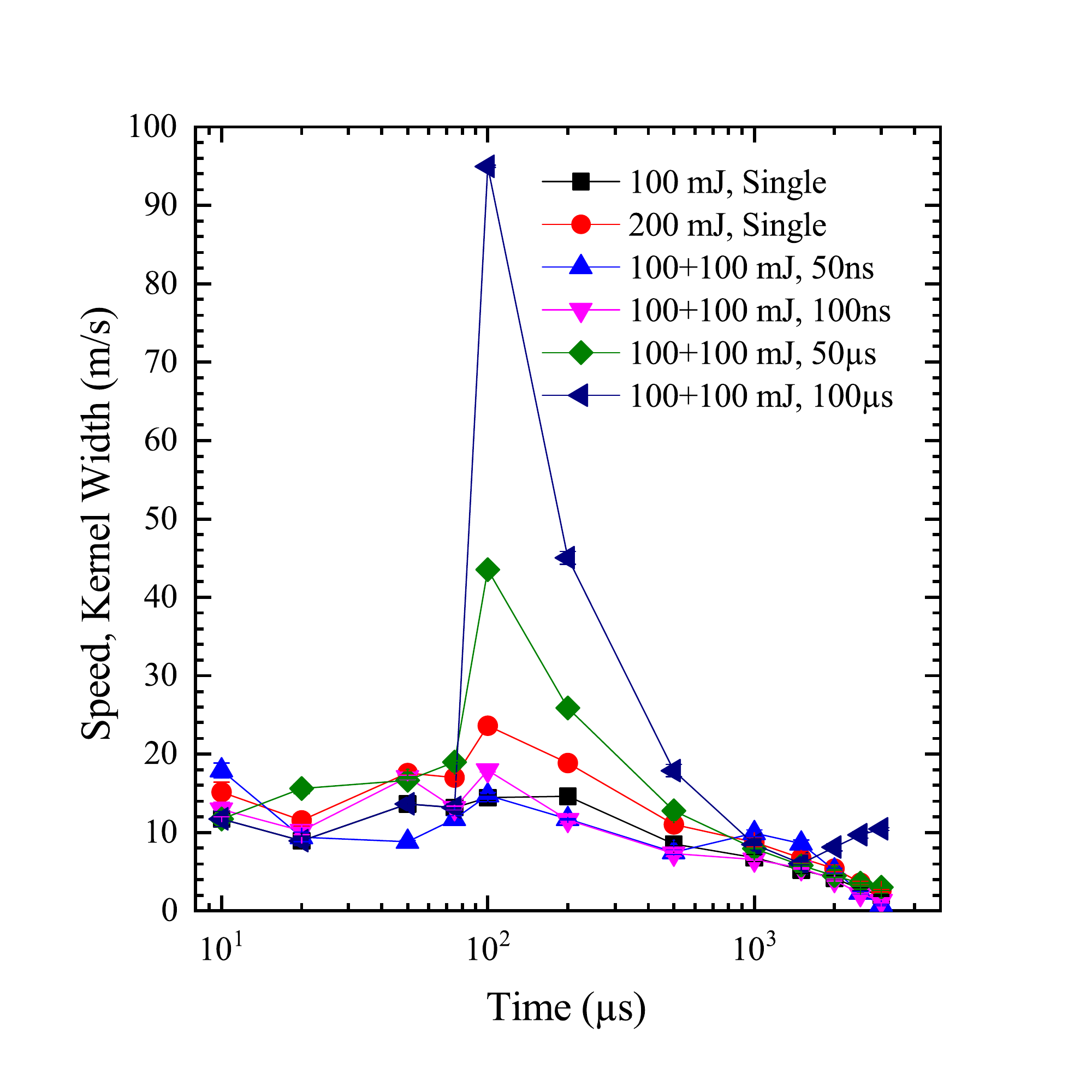}
            \caption[]%
            {{\small }}    
            \label{fig:speedwidth}
        \end{subfigure}
        \caption[The evolution of the plasma kernel ]
        {\small The evolution of the plasma kernel (hot plume) along the (a) height, and (b) width (between 10 $\mu{s}$ to 3 ms after the energy deposition) created by single-pulse LIB (with incident energies of 100 mJ and 200 mJ) and double-pulse LED (with incident energies of 100+100 mJ and the pulse interval $\Delta{t_{p}}$ = 50 $ns$, 100 $ns$, 50 $\mu{s}$, 100 $\mu{s}$) in quiescent air. The measured width and height are the maximum of the plasma kernel at any instant, the approach of the measurement is shown in Fig.~\ref{fig:normalization}.}
        \label{fig:width_height}
    \end{figure*}
    
    \begin{figure}[ht!]
\centering
\includegraphics[trim={0bp 0bp 0bp 0bp}, clip, width=1\columnwidth]{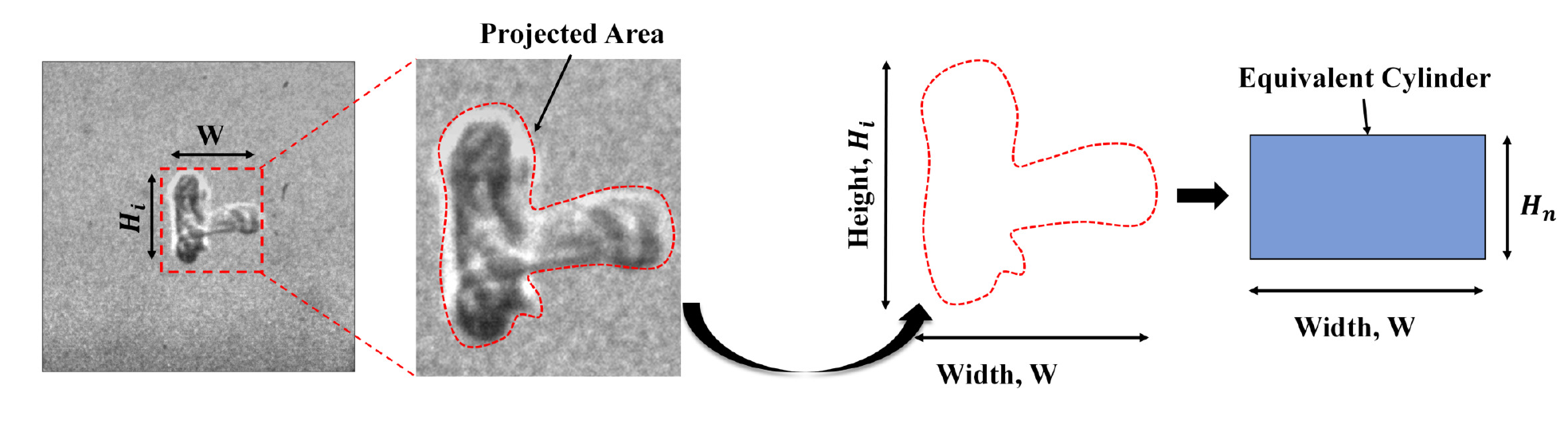} 
\caption{\label{fig:normalization}\small The approach for the normalization of original height of kernel to normalized average height, which is equivalent to diameter of cylinder. The projected area is the side view. The projection in the top view is exactly similar to the side view due to the symmetry of evolution. In order to estimate the volumetric expansion of the plasma core, the original height of the core can therefore be normalized to the equivalent diameter of the cylinder.
}
\end{figure}
\begin{figure*}[ht!]
        \centering
        \begin{subfigure}[b]{0.475\textwidth}
            \centering
            \includegraphics[trim={50bp 20bp 56bp 56bp}, width=\textwidth]{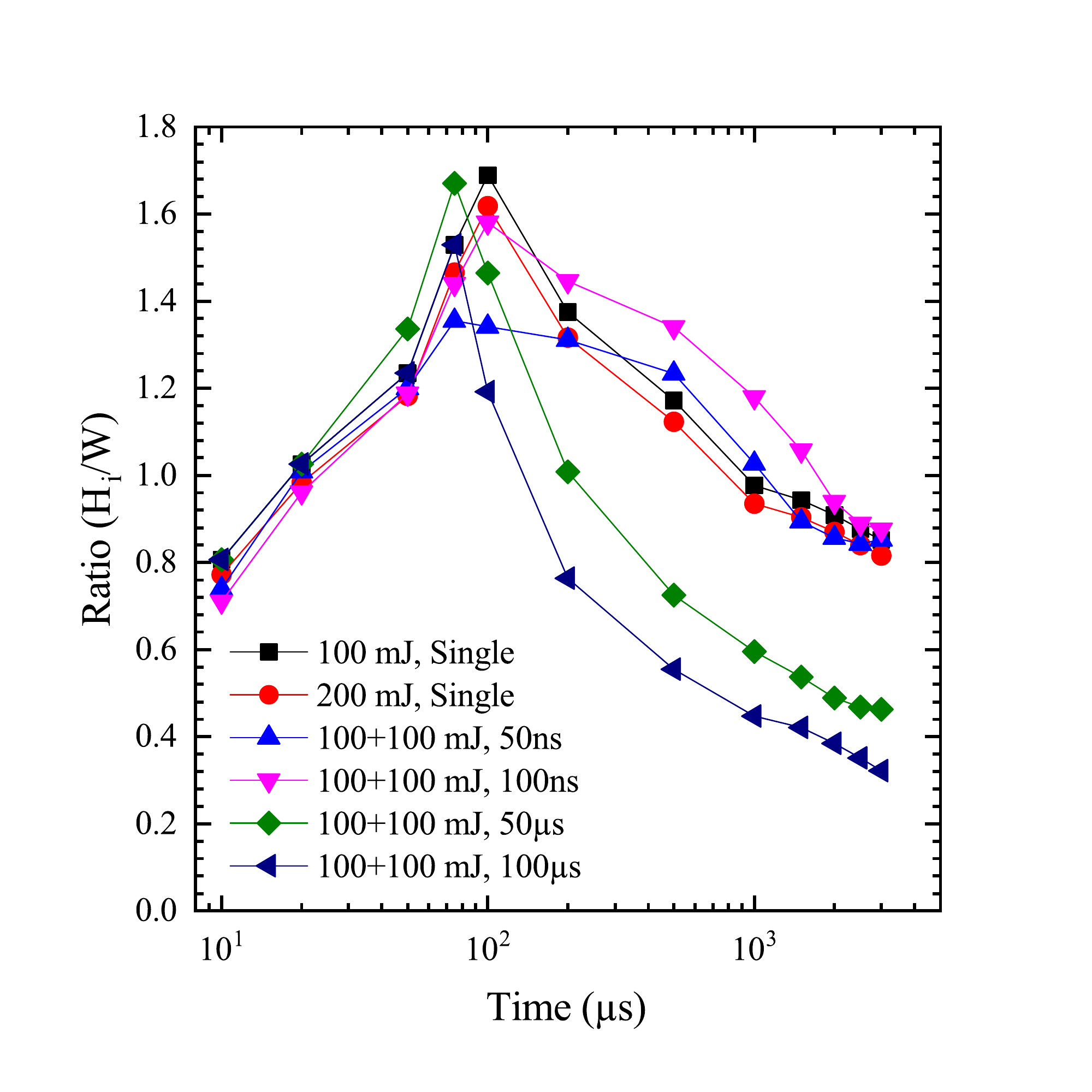}
            \caption[a]%
            {{\small }}    
            \label{fig:h_w_ratio}
        \end{subfigure}
        \hfill
        \begin{subfigure}[b]{0.475\textwidth}  
            \centering 
            \includegraphics[trim={50bp 20bp 56bp 56bp}, width=\textwidth]{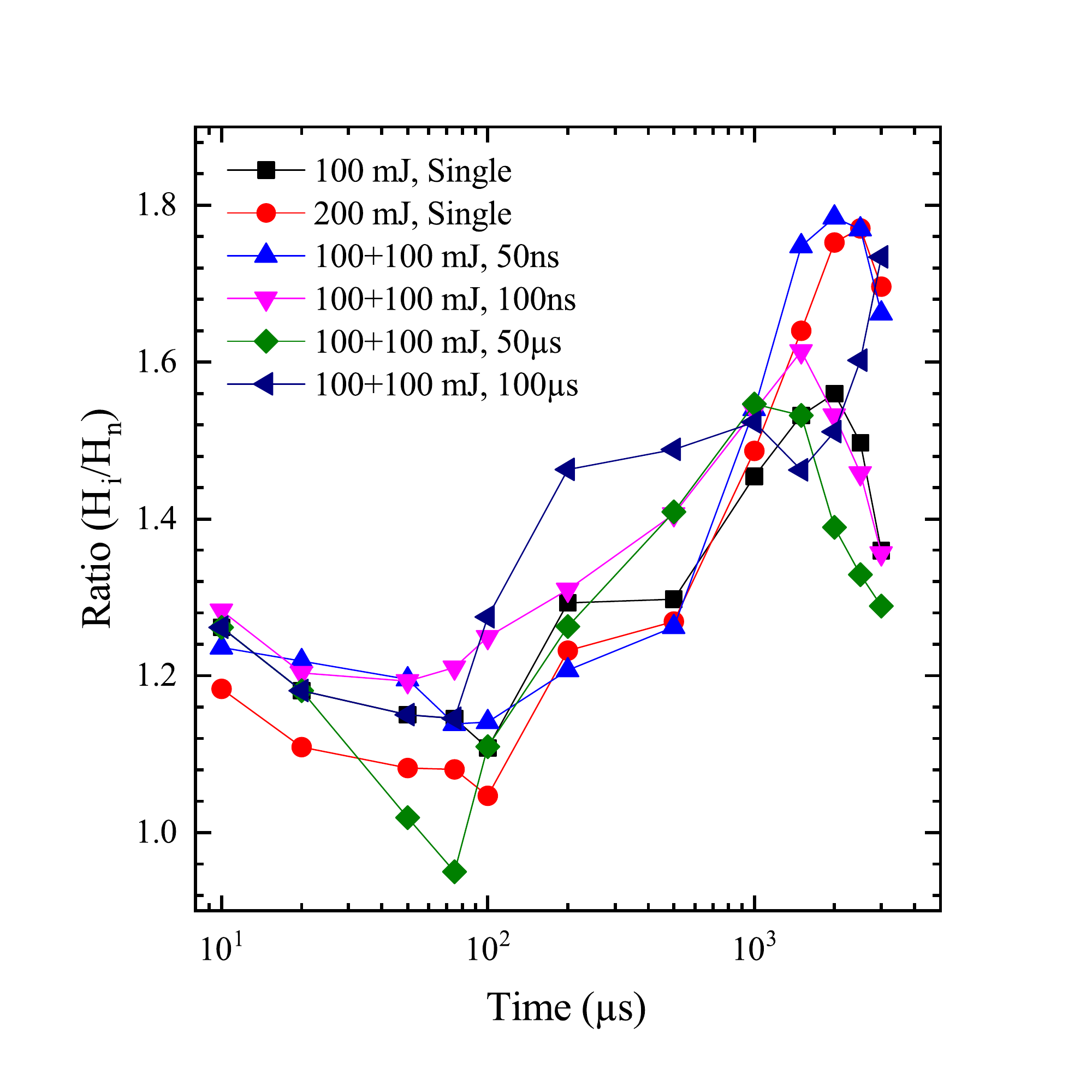}
            \caption[]%
            {{\small }}    
            \label{fig:new_ratio}
        \end{subfigure}
        \caption[The ratio of]
        {\small The ratio of (a) original height to width, and (b) original height to the normalized height at time ranging between 10 $\mu{s}$ to 3 $ms$ after the energy deposition.} 
        \label{fig:fig11}
    \end{figure*}
    
After the SP-LIB and DP-LEDs, the behavior of the plasma kernel needs to be clearly understood for various applications such as spectroscopy, combustion, hydrodynamics, surface cleaning, engraving, printing, welding, machining as well as for various biomedical and cosmetic applications. Therefore, in this regard, three parameters of the plasma kernel (width, height, and projection area) are measured with an in-house MATLAB algorithm. These parameters are further used to calculate the expansion speed of the plasma kernel. The spatio-temporal development of the plasma kernel along the height and width is shown in figures~\ref{fig:height} and \ref{fig:width}, respectively. The height and width of the plasma core in SP-LIB with 200 mJ pulse energy are approximately 1.24 and 1.30 times greater than those of 100 mJ SP-LIB and follow the same trend. The double pulse cases, however, do not follow the trend similar to SP-LIB or other DP-LEDs (see Fig.~\ref{fig:height} and \ref{fig:width}). The height expansion of plasma kernel for DP-LED ($\Delta{t_{p}}$ = 50 $ns$, 100 $ns$) is higher than that of 200 mJ SP-LIB and remains higher up to $\sim$ 100 $\mu{s}$. Thereafter, the expansion in DP-LEDs slows down due to the formation of lobes (see Fig.~\ref{fig:height} and \ref{fig:h_w_ratio}). The expansion of width and height in case of DP-LED ($\Delta{t_{p}}$ = 50 $ns$, 100 $ns$) is almost similar and near to SP-LIB (see Fig. \ref{fig:h_w_ratio} and \ref{fig:new_ratio}). However, when the pulse interval is increased (after $\sim$ 30 $\mu{s}$), the kernel width starts increasing rapidly due to the generation of the third and fourth lobe (see Fig. \ref{fig:Two} and \ref{fig:h_w_ratio}).\par
 \begin{figure}[ht!]
\centering
\includegraphics[trim={38bp 10bp 56bp 56bp}, clip, width=0.475\columnwidth]{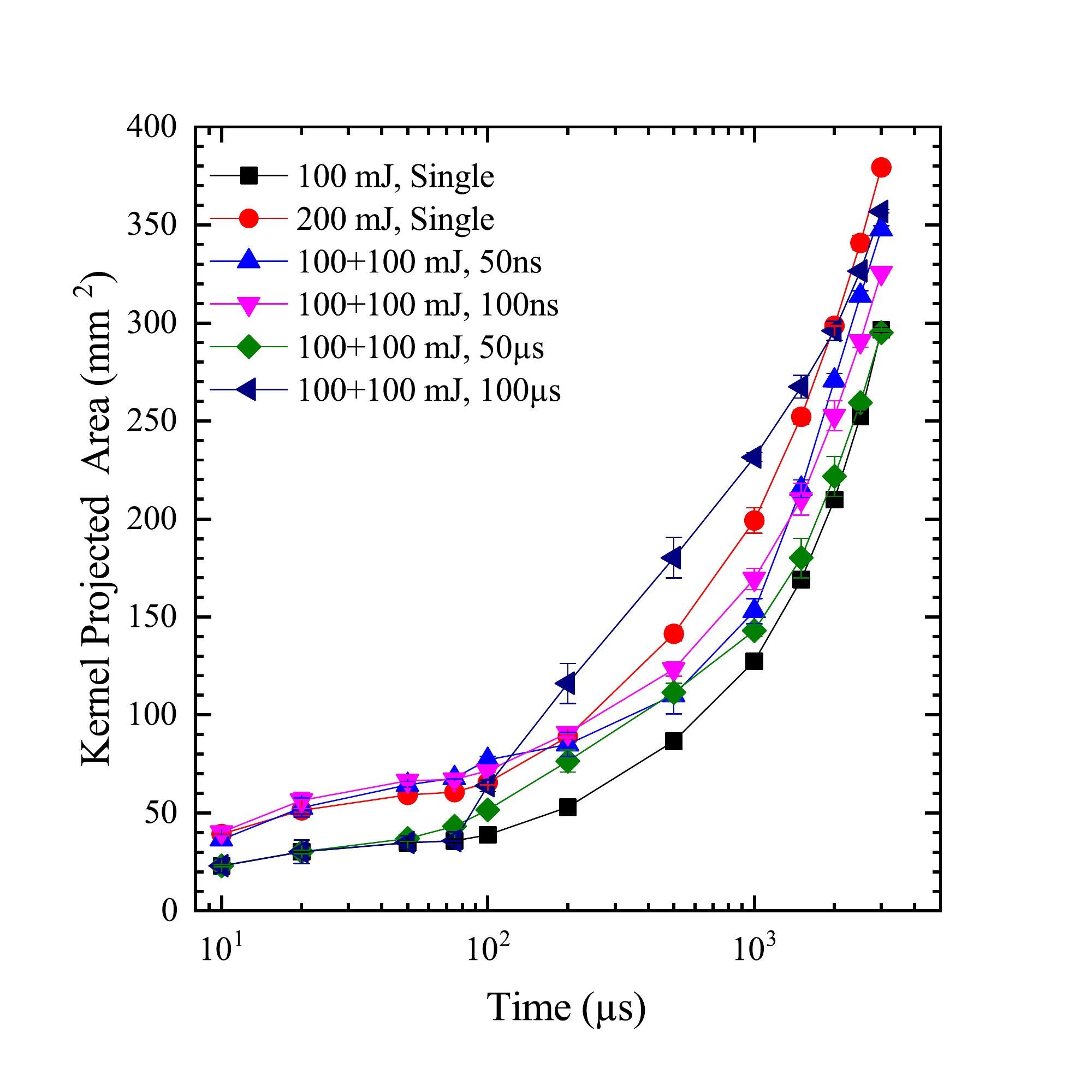} 
\caption{\label{fig:p_area}\small The projection area of the plasma kernel generated by SP-LIBs and DP-LEDs.}
\end{figure}
\begin{figure*}[ht!]
        \centering
        \begin{subfigure}[b]{0.475\textwidth}
            \centering
            \includegraphics[trim={50bp 20bp 56bp 56bp}, width=\textwidth]{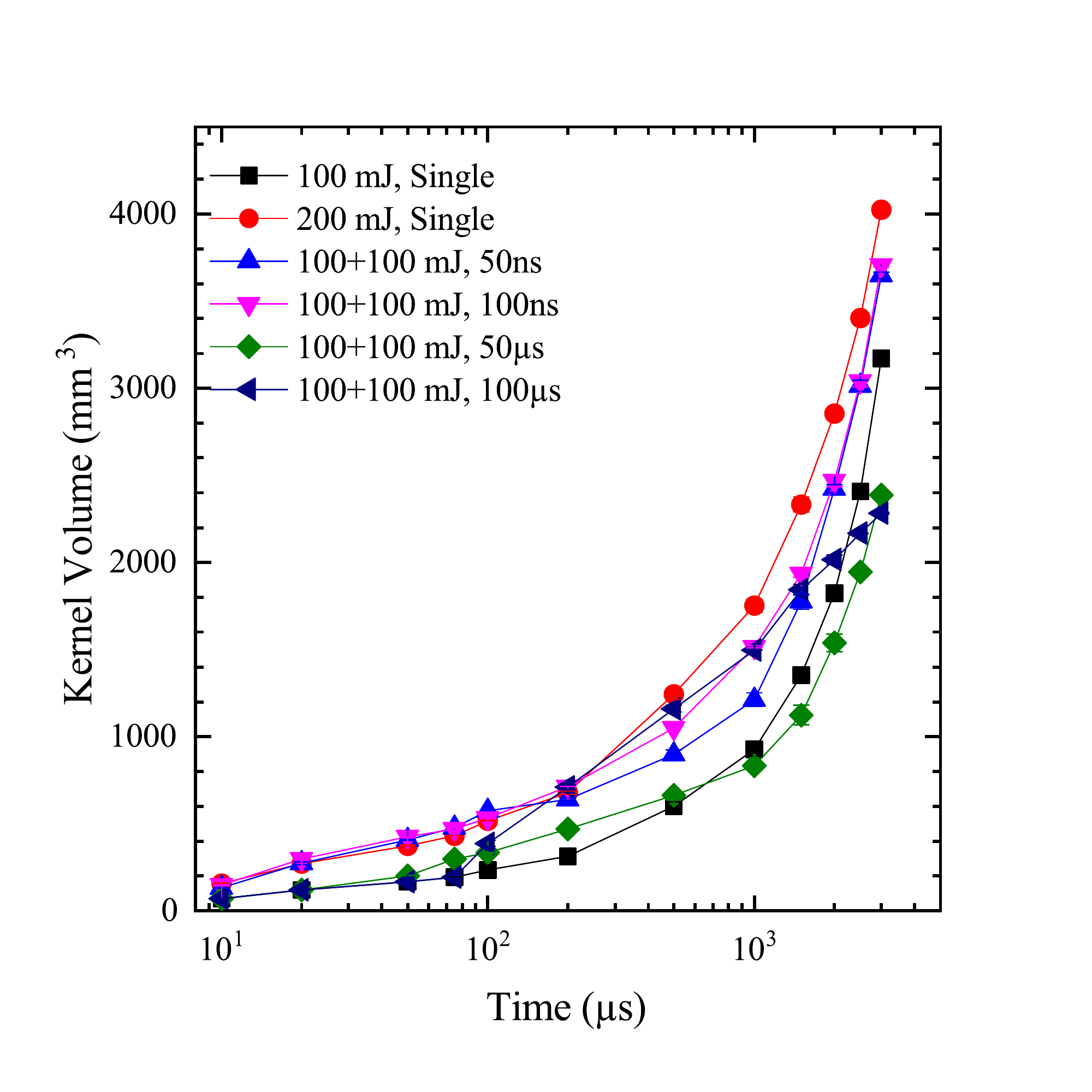}
            \caption[a]%
            {{\small }}    
            \label{fig:vol}
        \end{subfigure}
        \hfill
        \begin{subfigure}[b]{0.475\textwidth}  
            \centering 
            \includegraphics[trim={50bp 20bp 56bp 56bp}, width=\textwidth]{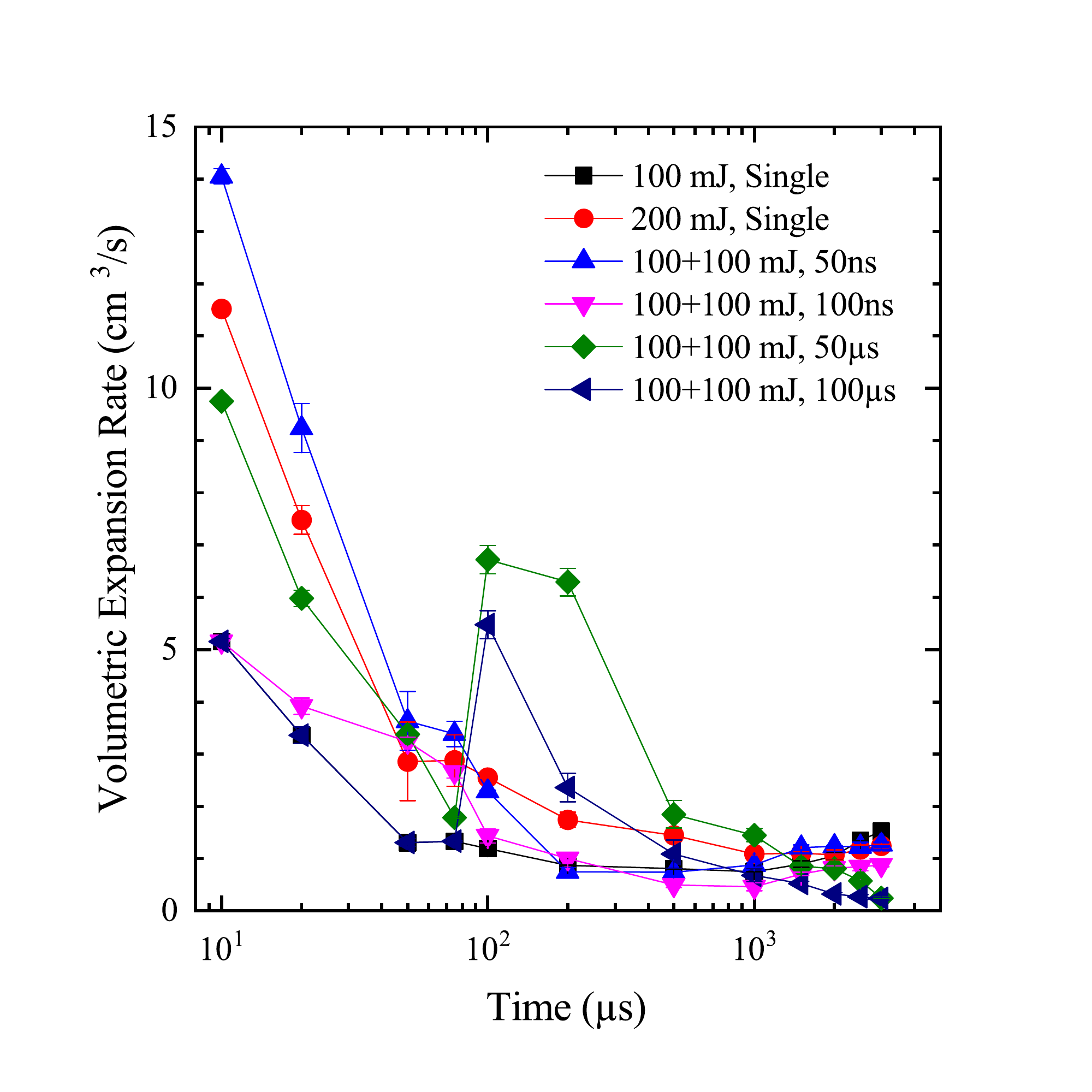}
            \caption[]%
            {{\small }}    
            \label{fig:ver}
        \end{subfigure}
        \caption[The ratio of]
        {\small The evolution of plasma kernel generated by SP-LIBs and DP-LEDs. (a) Volume of the plasma kernel, (b) Volumetric expansion rate. The volume of the kernel is calculated based on the approach given in Fig.~\ref{fig:normalization}.} 
        \label{fig:mean and std of nets}
    \end{figure*}

Figures ~\ref{fig:speedhight} and \ref{fig:speedwidth}  show the rate of expansion of the plasma kernel along with the height and width, which are very important in the applications mentioned above. The rate of expansion plays a crucial role in forming initial kernel (flame kernel in ignition), which later spreads and initiates the chemical reaction. The faster the speed, the faster the mixing, but in some cases, the width increases faster, and in some cases, it is the height. The ratio of height to width is shown in Fig.~\ref{fig:h_w_ratio}. However, it is not clear in which case the plasma kernel expands faster. Therefore, to understand which of the SP-LIBs and DP-LEDs are expanding faster, the volume and volumetric expansion rate (VER) of the plasma kernel are calculated.

 To calculate the volume and VER, the projected area of the plasma kernel is first determined using the MATLAB algorithm (Fig.~\ref{fig:p_area}). Based on this projection area, the original height of the plasma kernel is then normalized to a new height that corresponds to the diameter of the cylinder (Fig.~\ref{fig:normalization}). The relationship between the original height and normalized height is shown in Fig.~\ref{fig:new_ratio}. The normalized height is used to calculate the volume of the plasma kernel and later the VER over time (ranging between 10 $\mu{s}$ to 3 ms). The volume (Fig.~\ref{fig:vol}) and VER (Fig. \ref{fig:ver}) provides the actual evolution of the kernel, which is previously not clear from the height, width, and projection area. These two parameters (volume and VER) can provide the actual mixing rate of the plasma core with the surrounding gases or combustible mixtures. These measurements would help estimate the time it would take to initiate the post-energy deposition process (e.g. ignition) for various pulse interval and energy combinations.

 \newpage
 \subsection{Numerical simulation showing development of plamsa and shok-wave}
 
 The approximate values of physical properties (such as density, temperature, pressure, flow velocity, species concentration, etc.) during the laser-induced spark decay in quiescent atmospheric air are obtained from an in-house computer program. The Naiver-Stokes equations augmented by the species conservation equations are solved assuming that air obeys the perfect gas model. The simulation is performed in a Cartesian coordinate system in a two-dimensional framework considering thermal equilibrium and chemical non-equilibrium effects. The governing equations, numerical treatments and discretization techniques used in the present simulation are discussed by Padhi et al. in \cite{Padhi_2020}.
 
 The computation is performed in a square domain of 20 mm $\times$ 20 mm dimension. The grid convergence study for the geometry used was previously carried out in \cite{Joarder_2013}. The laser pulse is introduced instantaneously into the domain (the details of the numerical domain, laser discharge, and boundary conditions are discussed in \cite{Padhi_2020}). The physical properties after cessation of the laser pulse were determined using Helmholtz free energy minimization process \cite {joarder2017two}. The breakdown zone is characterized by the presence of ionized and dissociated species (O2, O, N2, N, NO, NO$^+$ and e$^-$) at high temperature and pressure. The simulation is carried out for SP-LIB (100 mJ ) and DP-LED (100 + 100 mJ) with a pulse interval of 50 $\mu{s}$. The characteristics time observed from the simulation for the chemical non-equilibrium (considering ionization, dissociation and recombination of species in the breakdown zone) ranges between 100-200 $ns$, while the thermal relaxation time is observed to be about 50-150 $ns$ for the range of energy considered in the present study. 
 
 \begin{figure}[ht!]
 	\centering
 	\includegraphics[trim={0bp 0bp 0bp 0bp},clip, width=1\columnwidth]{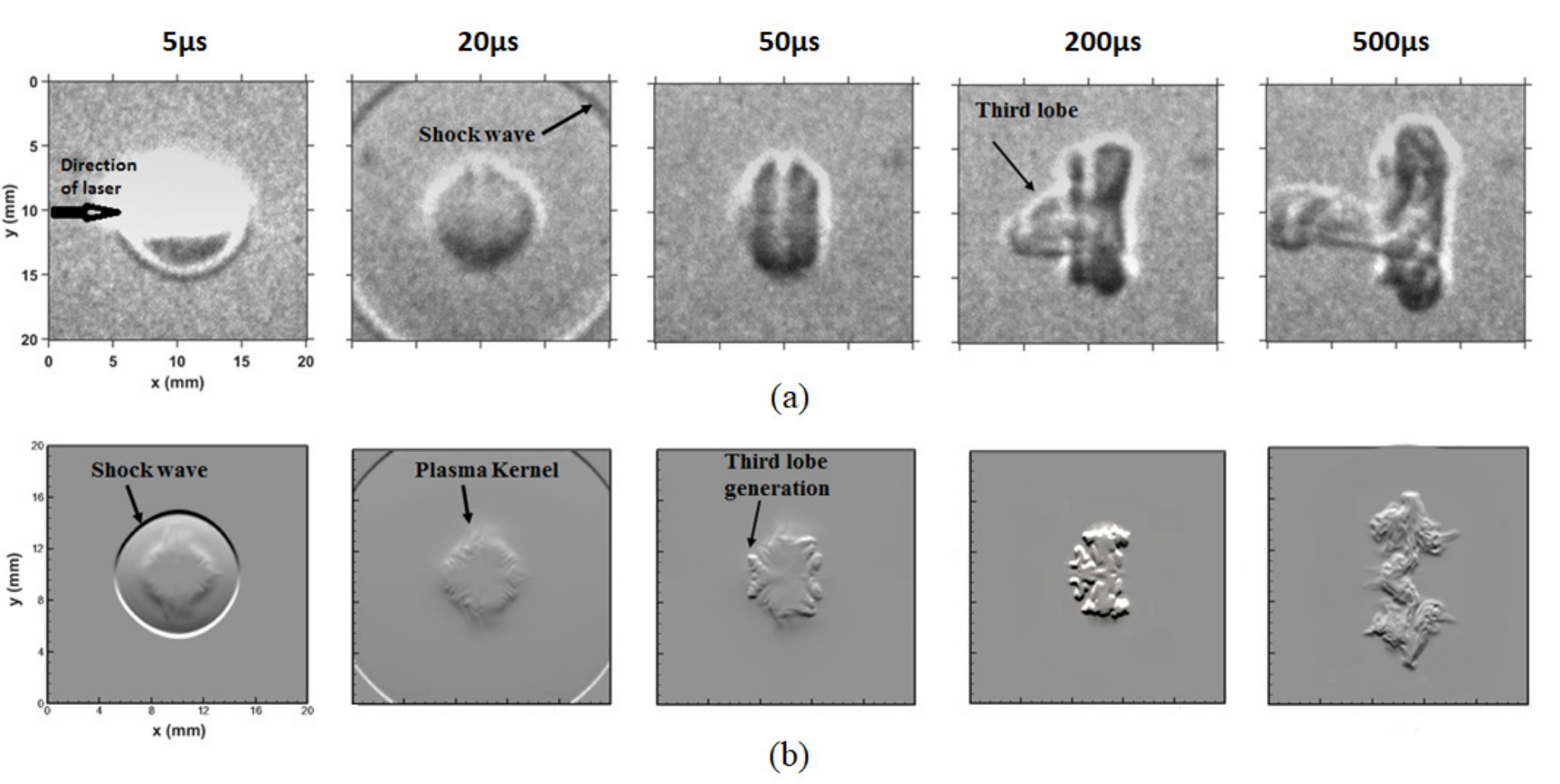}
 	\caption{\label{fig:single_pulse}\small  (a) Experimental (b) Numerical Schlieren images of shock-wave propagation in quiescent air of 100 mJ single pulse}
 \end{figure}

 \begin{figure}[ht!]
 	\centering
 	\includegraphics[trim={0bp 0bp 0bp 0bp},clip, width=1\columnwidth]{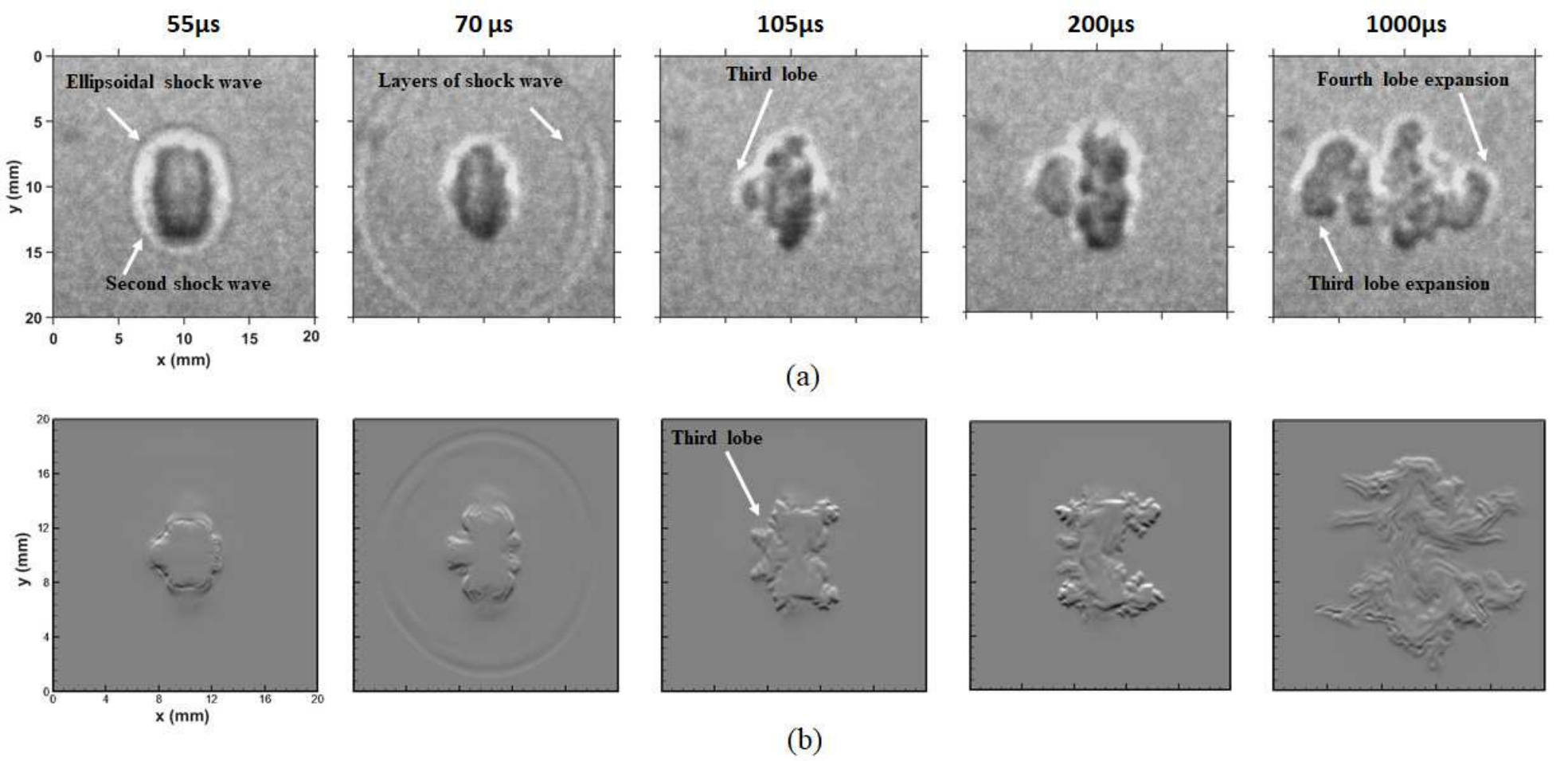}
 	\caption{\label{fig:double_pulse}\small Time sequential (a) Experimental (b) Numerical schlieren images of shock wave propagation in quiescent air for 100+100 mJ double pulse at $\Delta{t_{p}}$ =  50 $\mu{s}$ .}  
 	
 \end{figure}

 \begin{figure}[ht!]
 	\centering
 	\includegraphics[trim={0bp 0bp 0bp 0bp},clip, width=0.9\columnwidth]{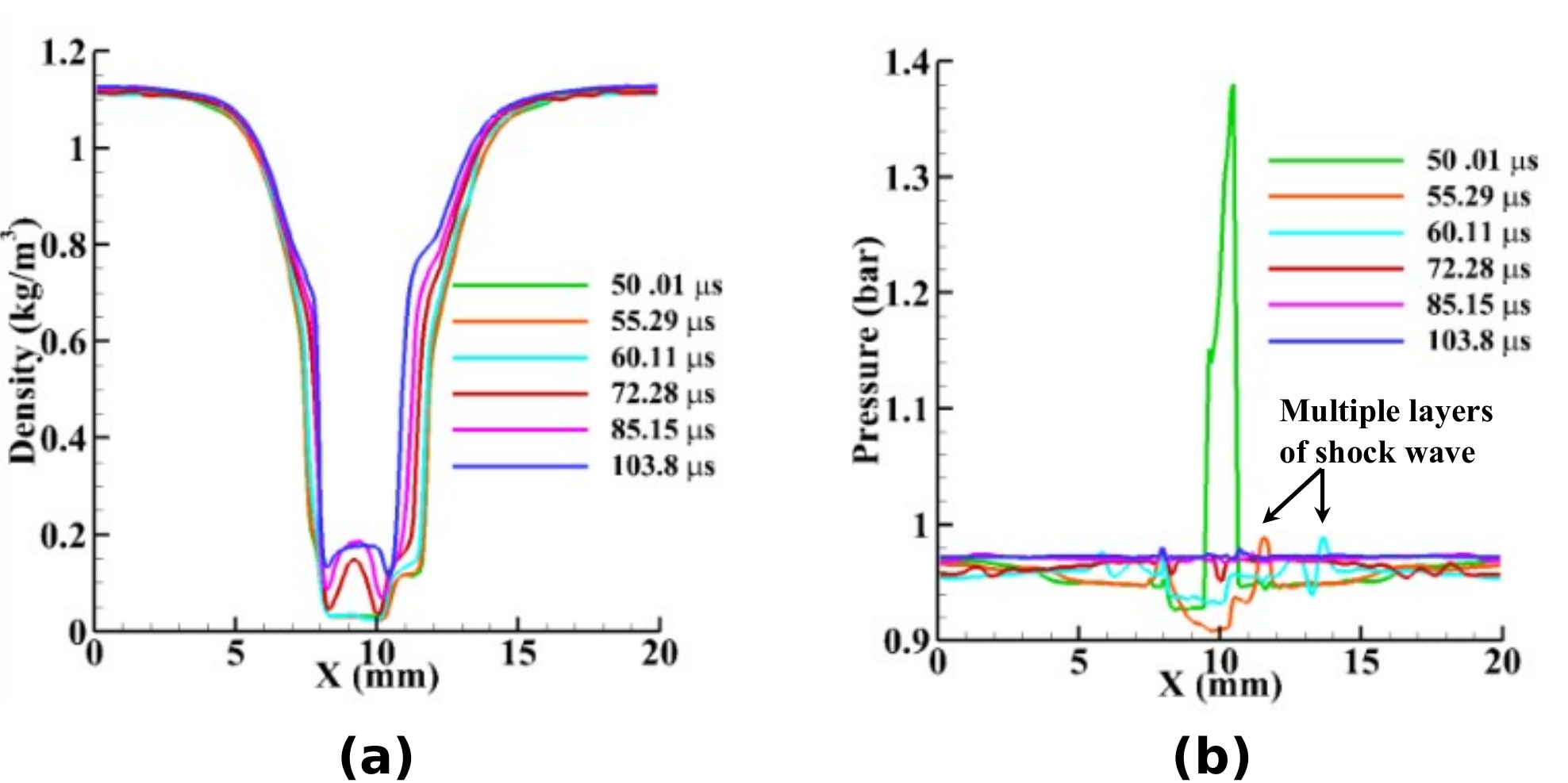}
 	\caption{\label{fig:den_press}\small The variation of (a) density, (b) pressure with successive double-pulse laser energy deposition. Multiple layers of shock waves and rarefaction waves develop from the core after the second pulse deposition. It can be seen that the strength of these multiple shock waves is relatively lower compared to that of a single pulse.}

\end{figure}

The time-sequential Schlieren images from the experiment and simulation are compared in Figure Fig.~\ref{fig:single_pulse} (a) and (b) for SP-LIB. It is illustrated that the plasma kernel's evolution and the propagation of shock waves follow a trend similar to the experiment. The trajectory of shock wave obtained from simulation is compared with the experimental result in \cite{Padhi_2020}, and they closely match with each other. The Schlieren images obtained from the simulation for a double pulse at 100 + 100 mJ at $\Delta{t_{p}}$ =  50 $\mu{s}$ are compared with experimental images (see Fig.~\ref{fig:double_pulse}). For DP-LED at $\Delta{t_{p}}$ =  50 $\mu{s}$, a new phenomenon of multiple shock wave propagation was observed both in the experiment and in the simulation. Multiple layers of ellipsoidal shock waves and rarefaction waves develop from the core, which later propagate as spherical waves (see Fig.~\ref{fig:double_pulse}). These shock waves originate from a low density and low-pressure core where species recombination is the dominating process ( see Fig.~\ref{fig:den_press} (a) and (b)). The strength of these shock waves is relatively small compared to a first pulse shock wave for the same total energy. In later times, the plasma at the core expands in either direction forming two lobes (third lobe and fourth lobe) as seen from Fig.~\ref{fig:double_pulse} at 1000 $\mu{s}$. The third lobe is observed opposite to the direction of the laser beam, and the fourth lobe is observed towards the direction of laser beam propagation. The probable reasons are already discussed in the earlier section.  However, the numerical simulation closely matches the experiment and supports our previous discussion on generating both the lobes and multiple shock waves.

\section{Conclusion}

High-speed Schlieren imaging and energy absorption measurements are performed during laser energy deposition in the quiescent air using two consecutive laser pulses. It has been observed that only the first pulse creates the breakdown, while the second pulse is absorbed by the plasma generated by the first breakdown. However, it is found that the second pulse induced a breakdown once the plasma created by the first pulse had cooled significantly, i.e. for $\Delta{t_{p}}$ $\geq$ 20 $\mu{s}$. The absorption from the second pulse decreases drastically between 200 $ns$ and 20 $\mu{s}$. This study uses a new approach to describe a previously unexplained (generation of third-lobe) and the newly observed physical phenomena (generation of fourth-lobe and multiple shock-waves). This new approach suggests that the third lobe in SP-LIBs is created due to the force against the radiation pressure that propagates against the direction of the laser beam. The fourth lobe and multiple shock waves are generated after the second-pulse deposition. The fourth lobe is generated in the laser beam's direction due to the radiation force on an optically thick compressed plasma kernel that is generated after the first pulse breakdown. The multiple shock waves, which propagated radially outward, are generated because the second pulse interacted and caused breakdown at different points within the perturbed density field of the plasma kernel. The similar phenomena was captured when we performed numerical simulation and compared with experiment. The volume and volumetric rate of expansion are calculated to determine the actual rate of expansion of the plasma in different single and successive laser energy deposition cases.

\section*{References}
\bibliographystyle{iopart-num}
\bibliography{sample.bib}

\end{document}